\documentclass[reprint,floatfix,twocolumn,superscriptaddress,showpacs, aps, prx,longbibliography]{revtex4-2} %Typical options above
\usepackage{graphicx} %insertion of eps figures
\usepackage{natbib} %implementation of \cite command; typical styles are apsrev is used for all the aps journals except rmp
\usepackage[usenames,dvipsnames]{color} % use colored text in latex
\usepackage{amsmath} % Defines extra environments for multiline displayed equations, as well as a number of other enhancements for math
\usepackage[urlcolor=blue, hyperindex, colorlinks, bookmarks=true,linkcolor=black,citecolor=black]{hyperref} % provides LaTeX the ability to create hyperlinks within the document
\usepackage{dcolumn}
\usepackage{amssymb} % {bm}% bold math
\usepackage{soul} %for underlining compatible with text wrapping; also other things
\usepackage{ifthen} %allows defining if then function
\usepackage{bbm}
\usepackage{comment}
\usepackage[ampersand]{easylist}
\usepackage{upgreek}
\usepackage{lineno}
\newcommand{\bra}[1]{\langle{#1}|}
\newcommand{\ket}[1]{|{#1}\rangle}

\def\l{\left}
\def\r{\right}

\def\be#1\ee{\begin{equation}#1\end{equation}}
\def\ba#1\ea{\begin{align}#1\end{align}}
\def\bg#1\eg{\begin{gather}#1\end{gather}}
\def\t{\text}

\newcommand{\uwiqc}{Institute for Quantum Computing, University of Waterloo, Waterloo, ON, Canada, N2L 3G1}
\newcommand{\uwphys}{Department of Physics and Astronomy, University of Waterloo, Waterloo, ON, Canada, N2L 3G1}
\newcommand{\uwee}{Department of Electrical and Computer Engineering,
University of Waterloo, Waterloo, ON, Canada, N2L 3G1}
\newcommand{\ngc}{Northrop Grumman Corporation, Linthicum, Maryland 21090, USA}
\newcommand{\mitrl}{Research Laboratory of Electronics, Massachusetts Institute of Technology, Cambridge, Massachusetts 02139, USA}
\newcommand{\mitll}{Lincoln Laboratory, Massachusetts Institute of Technology, Lexington, Massachusetts, 02421, USA}
\newcommand{\uscisi}{University of Southern California - Information Sciences Institute, Arlington, VA, 22203, USA}
\newcommand{\usccqist}{Center for Quantum Information Science \& Technology, University of Southern California, Los Angeles, California 90089, USA}
\newcommand{\uscee}{Department of Electrical \& Computer Engineering, University of Southern California, Los Angeles, California 90089, USA}
\def\shownoteal{1} %make this flag zero if you want text under \note to be shown
\newcommand{\nal}[1]{\ifthenelse{\shownoteal=1}{\textcolor{red}{[[#1]]}}{}}
\def\shownoteay{1} %make this flag zero if you want text under \note to be shown
\newcommand{\nay}[1]{\ifthenelse{\shownoteay=1}{\textcolor{orange}{[[#1]]}}{}}
\def\showaddmat{1} %make this flag zero if you want text under \personote to be shown
\newcommand{\addmat}[1]{\ifthenelse{\showaddmat=1}{\textcolor{Gray}{[[#1]]}}{}}
\def\shownote{1} %make this flag zero if you want text under \note to be shown
\newcommand{\note}[1]{\ifthenelse{\shownote=1}{\textcolor{Red}{[[#1]]}}{}}

\begin{document}
%\linenumbers
\title{Dissipative Landau-Zener tunneling in the crossover regime from weak to strong environment coupling}

\author{X. Dai}
\thanks{These two authors contributed equally. \\Corresponding authors: x35dai@uwaterloo.ca, (current address: Department of Physics, ETH Zurich, CH-8093 Zurich, Switzerland); rtrappen@uwaterloo.ca}
\affiliation{\uwiqc}
\affiliation{\uwphys}
\author{R. Trappen}
\thanks{These two authors contributed equally. \\Corresponding authors: x35dai@uwaterloo.ca, (current address: Department of Physics, ETH Zurich, CH-8093 Zurich, Switzerland); rtrappen@uwaterloo.ca}
\affiliation{\uwiqc}
\affiliation{\uwphys}
\author{H. Chen}
\altaffiliation{Current address: Department of Physics, Harvard University, Cambridge, MA, 02138, USA}
\affiliation{\usccqist}
\affiliation{\uscee}
\author{D. Melanson}
\affiliation{\uwiqc}
\affiliation{\uwphys}
\author{M. A. Yurtalan}
\affiliation{\uwiqc}
\affiliation{\uwphys}
\affiliation{\uwee}
\author{D. M. Tennant}
\altaffiliation{Current address: Rigetti Computing, Berkeley, California, 94710, USA}
\affiliation{\uwiqc}
\affiliation{\uwphys}
\author{A. J. Martinez}
\affiliation{\uwiqc}
\affiliation{\uwphys}
\author{Y. Tang}
\altaffiliation{Current address: 1QB Information Technologies (1QBit), Vancouver, BC, Canada, V6E 4B1}
\affiliation{\uwiqc}
\affiliation{\uwphys}
\author{E. Mozgunov}
\affiliation{\uscisi}
\author{J. Gibson}
\affiliation{\ngc}
\affiliation{Department of Physics and Astronomy, Dartmouth College, Hanover, NH 03755, USA}
\author{J. A. Grover}
\affiliation{\ngc}
\affiliation{\mitrl}
\author{S. M. Disseler}
\affiliation{\ngc}
\affiliation{\mitll}
\author{J. I. Basham}
\affiliation{\ngc}
\affiliation{QuEra Computing Inc., 1284 Soldiers Field Road, Boston, MA, 02135, USA}
\author{S. Novikov}
\affiliation{\ngc}
\affiliation{Atlantic Quantum Corp., Cambridge, MA 02139, USA}
\author{R. Das}
\author{A. J. Melville}
\author{B. M. Niedzielski}
\author{C. F. Hirjibehedin}
\author{K. Serniak}
\author{S. J. Weber}
\author{J. L. Yoder}
\affiliation{\mitll}
\author{W. D. Oliver}
\affiliation{\mitrl}
\affiliation{\mitll}
\author{K. M. Zick}
\affiliation{\ngc}
\affiliation{\uscisi}
\author{D. A. Lidar}
\affiliation{\usccqist}
\affiliation{\uscee}
\affiliation{Department of Chemistry, University of Southern California, Los Angeles, California 90089, USA}
\affiliation{Department of Physics, University of Southern California, Los Angeles, California 90089, USA}
\author{A. Lupascu}
\thanks{Corresponding author: adrian.lupascu@uwaterloo.ca}
\affiliation{\uwiqc}
\affiliation{\uwphys}
\affiliation{Waterloo Institute for Nanotechnology, University of Waterloo, Waterloo, ON, Canada N2L 3G1}

\date{ \today}
\begin{abstract}
Landau-Zener tunneling, which describes the transition in a two-level system during a sweep through an anti-crossing, is a model applicable to a wide range of physical phenomena. Realistic quantum systems are affected by dissipation due to coupling to their environments. An important aspect of understanding such open quantum systems is the relative energy scales of the system itself and the system-environment coupling, which distinguishes the weak- and strong-coupling regimes. Using a tunable superconducting flux qubit, we observe the crossover from weak to strong coupling to the environment in Landau-Zener tunneling. Our results confirm previous theoretical studies of dissipative Landau-Zener tunneling in the weak and strong coupling limits. We devise a spin bath model that effectively captures the crossover regime. This work is relevant for understanding the role of dissipation in quantum annealing, where the system is expected to go through a cascade of Landau-Zener transitions before reaching the target state.
\end{abstract}

\maketitle
\section{Introduction}
Landau-Zener (LZ) tunneling~\cite{landau_1932_zurtheorieenergieubertragung,zener_1932_nonadiabaticcrossingenergy,majorana_1932_atomiorientaticampo,stuckelberg_1932_theoryinelasticcollisions} describes non-adiabatic transitions through an anti-crossing in a two-state quantum system with a linearly changing energy separation. The LZ model is applicable to a wide range of physical phenomena, such as atomic collisions~\cite{nikitin_1984_theoryslowatomic}, chemical reactions ~\cite{hanggi_1990_reactionratetheoryfifty}, and molecular magnets~\cite{leuenberger_2003_quantumspindynamics}. Physical realizations of LZ tunneling are influenced by system-environment coupling~\cite{oliver_2005_machzehnderinterferometrystrongly,zenesini_2009_timeresolvedmeasurementlandauzener,troiani_2017_landauzenertransitioncontinuously,zhang_2018_symmetrybreakingassistedlandauzener,zhu_2021_crossoveradiabaticnonadiabatic}, and there are extensive theoretical studies on the effects of dissipation on LZ tunneling~\cite{kayanuma_1984_nonadiabatictransitionslevel,ao_1989_influencedissipationlandauzener,kayanuma_1998_nonadiabatictransitionlevel,wubs_2006_gaugingquantumheat,saito_2007_dissipativelandauzenertransitions,amin_2008_thermallyassistedadiabatic,nalbach_2009_landauzenertransitionsdissipative,arceci_2017_dissipativelandauzenerproblem,wang_2021_nonadiabaticevolutionthermodynamicsa}. 

Studying the effects of dissipation in quantum systems is of both fundamental interest~\cite{weiss_2012_quantumdissipativesystems,breuer_2007_theoryopenquantum} and important for practical applications such as quantum information processing~\cite{nielsen_2010_quantumcomputationquantum}. Dissipation is of particular concern to analog quantum computation, where error due to dissipation cannot be indefinitely suppressed due to the lack of a fault-tolerance threshold~\cite{nielsen_2010_quantumcomputationquantum}. One prominent analog quantum algorithm is quantum annealing~\cite{das_2008_colloquiumquantumannealing,albash_2018_adiabaticquantumcomputation,hauke_2020_perspectivesquantumannealinga}, where a system goes through one or multiple LZ tunnelings to reach some many-body quantum state of interest, relevant to quantum simulation or optimization tasks. The role of dissipation 
in quantum annealing is still largely an open question, with previous studies suggesting that dissipation could either improve the annealing performance by cooling the system~\cite{dickson_2013_thermallyassistedquantum,marshall_2019_powerpausingadvancing,chen_2020_whywhenpausing} or hamper the tunneling process~\cite{denchev_2016_whatcomputationalvalue,boixo_2016_computationalmultiqubittunnelling,bando_2022_breakdownweakcouplinglimit}, depending on the relative strength between the system-environment coupling and relevant system energy scales. 

For the two-state LZ tunneling with a single sweep across the anti-crossing, analytical solutions for the transition probabilities in the closed-system case were obtained in Refs.~\cite{landau_1932_zurtheorieenergieubertragung,zener_1932_nonadiabaticcrossingenergy,majorana_1932_atomiorientaticampo,stuckelberg_1932_theoryinelasticcollisions}. Specifically, for a given tunneling amplitude (or equivalently the minimum spectral gap size) $\Delta$ and sweep velocity $v$, the probability for the system to tunnel to the different diabatic state is $1-P_\text{LZ}$, where $P_\text{LZ}$ in the coherent limit is given by
\begin{align}
P_{\text{LZ}} = \exp{\left(-\frac{\pi\Delta^2}{2\hbar v}\right)}\label{eq:PLZ}.
\end{align}

For most devices made for quantum information processing tasks, dissipation is well described by assuming weak coupling to a Markovian environment. Applying these assumptions to the Landau-Zener problem and considering noise coupled longitudinally (or equivalently to the diabatic states), it has been found that the dependence of the transition probability on sweep rate is unchanged if the noise temperature is low compared to the system's minimum spectral gap at the anti-crossing, and approaches the asymptotic value of $1/2$ when the noise temperature is high~\cite{ao_1989_influencedissipationlandauzener}. At intermediate temperatures, it has been predicted that the noise leads to non-monotonic dependence of the tunneling probability on the sweep rate, due to the competition between adiabaticity, thermal excitation near the anti-crossing, and thermal relaxation after the anti-crossing~\cite{nalbach_2009_landauzenertransitionsdissipative}. This non-monotonic dependence has not yet been observed, due to the intricate requirement on the various energy scales. 

Strong system-environment coupling occurs both naturally in physical phenomena such as electron transfer~\cite{lambert_2013_quantumbiology}, and in engineered quantum devices such as quantum annealers~\cite{smirnov_2018_theoryopenquantuma}. A general description in the strong coupling regime is challenging due to finite system-bath correlations and non-Markovian effects. Here we focus on strong low-frequency noise, which is ubiquitous in solid-state systems such as superconducting qubits~\cite{dutta_1981_lowfrequencyfluctuationssolids,paladino_2014_noiseimplicationssolidstate}. Superconducting flux qubits are one of the leading platforms for quantum annealing, and the dominant noise is the intrinsic flux noise~\cite{yoshihara_2006_decoherencefluxqubits}, the physical origin of which remains elusive to date. While the noise power spectral density of flux noise has been well-characterized~\cite{bylander_2011_noisespectroscopydynamical,quintana_2017_observationclassicalquantumcrossover,rower_2023_evolutionfluxnoise}, a general open-system model for flux noise remains challenging due to its long correlation time. In the specific case of quantum tunneling with small tunneling amplitude, it has been shown in macroscopic resonant tunneling (MRT) experiments that longitudinally coupled low-frequency noise can be modeled as an environmental polarization that preferentially aligns with the diabatic states of the system~\cite{amin_2008_macroscopicresonanttunneling,harris_2008_probingnoiseflux}, which can be formally described using the polaron transformation~\cite{boixo_2016_computationalmultiqubittunnelling,chen_2022_hamiltonianopenquantum}. For Landau-Zener tunneling, the transition probability in the presence of strong low-frequency noise remains unchanged from the coherent case, provided that the integrated noise amplitude is higher than the noise temperature~\cite{johansson_2009_landauzenertransitionssuperconducting}.

In this work, we experimentally investigate LZ tunneling with a tunable flux qubit for a wide range of minimum gap $\Delta$ and sweep velocity $v$. We present detailed numerical modeling of the open-system effects, which goes beyond the phenomenological models adopted in previous experiments~\cite{oliver_2005_machzehnderinterferometrystrongly,quintana_2013_cavitymediatedentanglementgeneration,troiani_2017_landauzenertransitioncontinuously}. Comparing the experimental and numerical models reveals that as the tunneling amplitude $\Delta$ decreases, the dominant open-system effects display a crossover from that of a weakly-coupled Markovian environment to a strongly coupled non-Markovian environment. We further devise a toy environment model in terms of a spin bath, which shows qualitative agreement with the experiment for the full range of parameters. The spin bath model suggests that the crossover arises as the time scale for the qubit to complete the tunneling, determined by $\Delta$, becomes comparable to the time scale over which the low-frequency environment reorganizes itself into the lower energy configuration. Our experiment probes a previously unexplored regime of system-environment coupling and the theoretical analysis presents new directions toward understanding and modeling intrinsic flux noise in superconducting circuits, which is both of fundamental importance and practical value for assessing the capability of analog quantum processors such as the quantum annealer.

\begin{figure}[ht]
    \centering
    \includegraphics{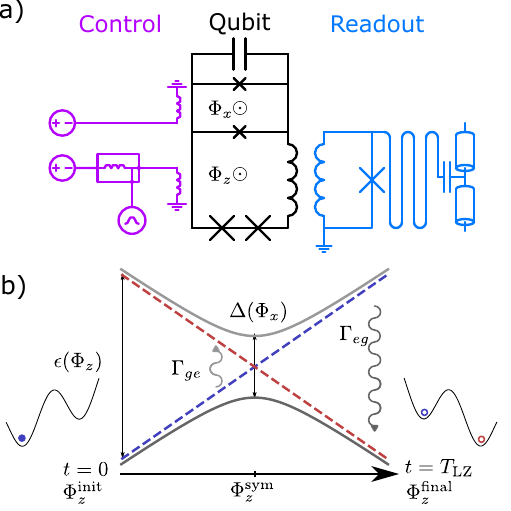}
    \caption{The flux qubit and the dissipative LZ transition. (a) Schematic of the tunable capacitively-shunted flux qubit and the control and readout circuitry. The flux biases $\Phi_x, \Phi_z$ are each supplied by a DC voltage source and the $\Phi_z$ is further controlled by a fast arbitrary waveform generator, joined to the DC control through a diplexer. Readout is done by measuring the transmission through an rf-SQUID terminated waveguide resonator coupled inductively to the qubit. (b) Schematic representation of the LZ sequence. The blue and red dashed lines indicate the energies of the diabatic states, which are separated by $\epsilon(\Phi_z)=2I_p(\Phi_z-\Phi_z^\text{sym})$. The grey lines indicate the eigenenergies of the qubit, which has a minimum gap of $\Delta$ at the symmetry point $\Phi_z^\text{sym}$. The curly arrows indicate the dominant open-system effects in the LZ measurements in the weak-coupling limit, which are excitations around the symmetry point and relaxation after the symmetry point. The double-well plots on either side of the energy level diagram are a representation of the qubit potential at the beginning and end of the LZ sweep.}
    \label{fig:SchematicAndPulse}
\end{figure}

\section{Results}
\subsection{The tunable capacitively-shunted flux qubit}
Our experiments are performed using a two-level quantum system implemented using a tunable superconducting capacitively-shunted flux qubit~\cite{yan_2016_fluxqubitrevisited, novikov_2018_exploringmorecoherentquantum, khezri_2021_annealpathcorrectionflux}. A schematic of the experiment setup is shown in Figure~\ref{fig:SchematicAndPulse}(a). The qubit circuit consists of two flux loops, designated as $z, x$ respectively, including Josephson tunnel junctions. Under suitable flux bias conditions, the circuit has a double-well potential energy landscape, with the two wells corresponding to persistent current flowing in opposite directions in the $z$ loop. When cooled down to the base temperature of a dilution fridge, the system is confined to the respective ground states of the two potential wells, described by the two-state (qubit) Hamiltonian
\begin{align}
    H_q &= -\frac{\epsilon(\Phi_z)}{2}\sigma_z - \frac{\Delta(\Phi_x)}{2} \sigma_x,\label{eq:QubitHam}
\end{align}
with $\sigma_{z,x}$ being the Pauli operators. Here $\epsilon=2I_p(\Phi_z-\Phi_z^{\text{sym}})$ and $\Delta$ are respectively the bias and tunnelling amplitude between persistent current states, $I_p$ is the persistent current, $\Phi_{z(x)}$ is the flux bias in the $z(x)$ loop, and $\Phi_z^\text{sym}$ is the $\Phi_z$ bias which gives a symmetric double-well potential. The $\Phi_x, \Phi_z$ biases are controlled by DC voltage sources and the $\Phi_z$ bias is additionally coupled to a fast arbitrary waveform generator (AWG), combined with the DC control through a diplexer. Readout of persistent current states is done by inductively coupling the qubit $z$ loop to a flux-sensitive resonator. The circuit is capacitively coupled to a waveguide used to send microwave signals, allowing resonant excitation of the circuit. A circuit network model is fit to the spectroscopically measured transition frequencies, which allows for determining the circuit parameters. The two-state model parameters $I_p$ and $\Delta$ can then be obtained from the circuit model at arbitrary flux biases near the symmetry point $\Phi_z = \Phi_z^\text{sym}$. Details of the experimental setup and calibration measurements are given in Supplementary Notes 1 to 6.

\subsection*{LZ tunneling in the short-time limit}
A diagrammatic representation for the LZ measurement is shown in Figure~\ref{fig:SchematicAndPulse}(b). At $t=0$, the qubit is prepared in the left well at $\Phi_z^\text{init}\approx-0.005\Phi_0+\Phi_z^\text{sym}$. Then the qubit $z$ flux $\Phi_z$ is linearly swept to $\Phi_z^\text{final}\approx0.005\Phi_0+\Phi_z^\text{sym}$ over the duration $T_{\text{LZ}}$. The initial and final values of $\Phi_z$ ensure the LZ sweep starts and ends far enough from the anti-crossing so that the qubit energy eigenstates approximately overlap with the persistent current states. The qubit state population after the sweep is read out by measuring the state-dependent transmission through the resonator. The sequence is repeated for a range of $\Phi_x$ and $T_{\text{LZ}}$ values. With decreasing $\Phi_x$, $\Delta$ decreases nearly exponentially whereas $I_p$ increases by about $10\%$ over the entire range.

\begin{figure}
    \includegraphics{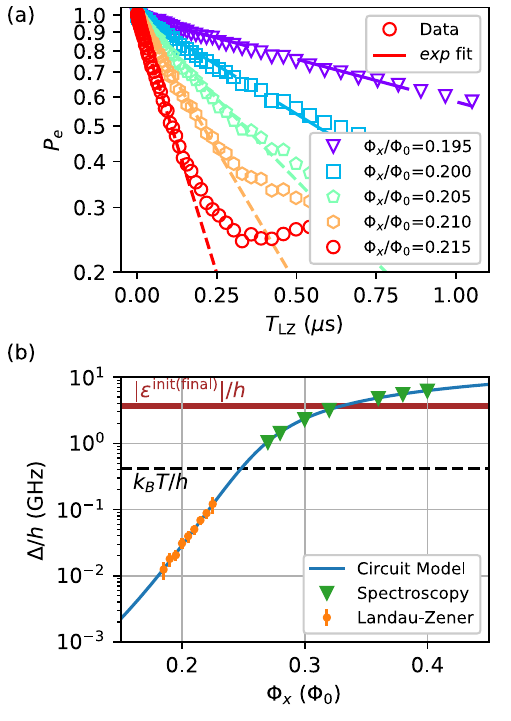}
    \caption{LZ data in the short sweep time range and fitted effective gap. (a) Experimental data (open markers) and exponential fits (dashed lines) to the final excited states $P_e$ versus the sweep time $T_{\text{LZ}}$. The maximum sweep time included in the fit is determined adaptively, by first starting from $30~\text{ns}$ and then increasing until the mean square loss of the fit exceeds $0.01$. (b) The minimum gap $\Delta$ versus x-bias flux $\Phi_x$ from spectroscopy (green triangle), LZ (orange dots), and circuit model (blue line). Error bars in the LZ data represent standard error propagated from the exponential decay fit error. The circuit model is a result of fitting spectroscopy data for a range of $\Phi_x$, $\Phi_z$ (not shown here). The black dashed line indicates the noise temperature, which is assumed to be close to the base temperature of the dilution fridge, $T=20\,\text{mK}$. The red horizontal band indicates the qubit energy splitting at the beginning and the end of the LZ sweep.}
        \label{fig:PeAndGap}
  %\end{minipage}
\end{figure}

The measured final excited state probabilities $P_e$ versus $T_{\text{LZ}}$ at short times are shown in Figure~\ref{fig:PeAndGap}(a). In the weak-coupling limit, the system is expected to behave nearly coherently for short sweep times, implying that the final excited state probabilities are well described by Eq.~\ref{eq:PLZ}, with the sweep velocity given by
\begin{equation}
v=\frac{2I_p(\Phi_z^\text{final}-\Phi_z^\text{init})}{T_{\text{LZ}}}.~\label{eq:SweepVelocity}
\end{equation}
To confirm the coherent-limit behavior, we fit an exponential decay to the short-time decay of the measured final excited state probabilities, and then convert the decay constant to an effective gap $\Delta_{\text{LZ}}$ assuming $I_p$ given by the circuit model. The extracted values of the effective gap $\Delta_{\text{LZ}}$ are compared to values of $\Delta$ given by the circuit model in Figure~\ref{fig:PeAndGap}(b). There is excellent agreement for the full range of $\Phi_x$ measured in LZ experiments, with $\Delta$ in the range of $12-120\,\text{MHz}$. Further increasing or decreasing $\Delta$ would prevent us from observing the full transition from non-adiabatic to adiabatic evolution within the range of sweep time afforded by our flux control bandwidth.

\subsection*{LZ tunneling in the long-time limit}
After confirming the short-time behavior, we observe the dynamics at longer sweep times, where coupling to the environment is expected to affect the LZ transition. We first discuss the characterization of the environment. Measurement of the noise spectrum is done based on its effect on qubit relaxation and dephasing at $\Delta/h\gtrsim 1~\text{GHz}$~\cite{trappen_2023_decoherencetunablecapacitively}, where the weak coupling limit holds. We find that the coherence is flux noise limited and can be explained by a noise power spectral density (PSD) consistent with previous work~\cite{yan_2016_fluxqubitrevisited,quintana_2017_observationclassicalquantumcrossover,lanting_2011_probinghighfrequencynoise}, where the noise power varies to a good approximation as $1/f$, with $f$ the frequency, up to $1~\text{GHz}$, combined with quasi-ohmic noise at higher frequencies. Our noise measurements, which are sensitive to the symmetrized noise power, combined with the assumption that the environment is in thermal equilibrium at the fridge base temperature, allow us to write the quantum noise PSD (unsymmetrized) as

\begin{align}
    S_{\lambda}(\omega) &= S_{\lambda,1/f}(\omega) + S_{\lambda, \text{ohmic}}(\omega),\\
    S_{\lambda, 1/f}(\omega) &= \frac{A_{\lambda}\omega}{|\omega|^\alpha}\l[1+\coth\l(\frac{\beta\hbar\omega}{2}\r)\r]~\text{and}\\
    S_{\lambda,\text{ohmic}}(\omega)&=B_{\lambda}\omega|\omega|^{\gamma-1}\l[1+\coth\l(\frac{\beta\hbar\omega}{2}\r)\r],
\end{align}
with $\lambda\in\{\Phi_x, \Phi_z\}$. Here $\beta=1/k_BT$ is the inverse temperature and $A_{\lambda}(B_{\lambda})$ and $\alpha(\gamma)$ characterize the amplitude and frequency dependence of the $1/f$(quasi-ohmic) component respectively. Given the smaller noise power and coupling matrix elements of the $\Phi_x$ noise for flux biases probed in the LZ measurement, we only consider $\Phi_z$ noise from here onward~(see Supplementary Note 7). 

\begin{figure*}
\centering
\includegraphics[]{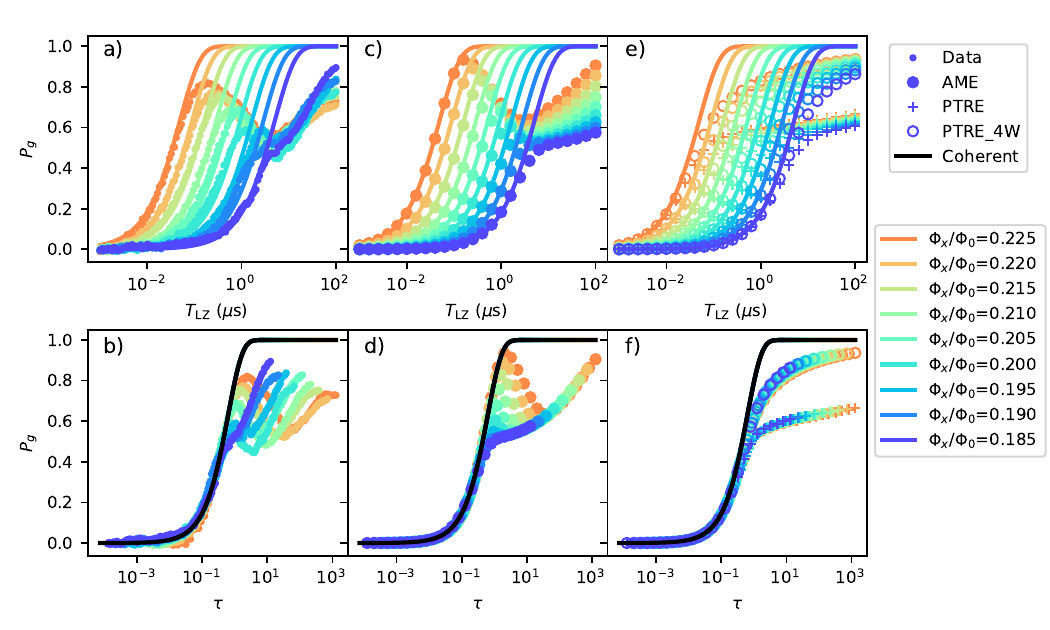}
\caption{LZ data for the full range of $\Phi_x$ and sweep time, and comparison with simulation. Final ground-state probability $P_g$ versus the sweep time $T_\text{LZ}$ (top) and the dimensionless sweep time $\tau=\Delta^2/\hbar v$ (bottom), for (a, b) experimental data, (c,d) Adiabatic master equation simulation (labeled as AME) results with nominal noise parameters and (e,f) Polaron-transformed Redfield equation simulation with nominal noise parameters (labeled as PTRE) and 4 times larger MRT width $W$ (labeled as PTRE\_4W). All panels also contain the coherent limit given by $P_g=1-P_{\text{LZ}}$.}
\label{fig:ExpAMEPTRE}
\end{figure*}

The measured final ground state probabilities $P_g=1-P_e$ for different $\Phi_x$ are shown in Figure~\ref{fig:ExpAMEPTRE}(a,b), versus the full range of sweep time $T_{\text{LZ}}$ and the dimensionless time $\tau=\Delta^2/\hbar v$, with $\Delta$ being the predicted value from the circuit model, and $v$ the sweep velocity defined in Eq.~\ref{eq:SweepVelocity}. Analyzing the dependence on both the actual time $T_{\text{LZ}}$, and dimensionless time $\tau$ allows us to make complementary observations about the changes in the effect of the environment as $\Phi_x$, or equivalently $\Delta$, is tuned. 

\subsection*{Simulation with master equations in the weak- and strong-coupling limit}
The weak-coupling limit between the system and the environment is expected to apply when the system-environment coupling strength is much smaller than the system's own energy scale, in our case $\Delta$. In the weak coupling limit, the adiabatic master equation (AME)~\cite{albash_2012_quantumadiabaticmarkovian} can be applied, where the environmental effect is assumed to be Markovian and leads to thermal transitions and decoherence between the energy eigenstates of the system. The AME-simulated final ground state probabilities are shown in Figure 3(c,d). For large $\Phi_x$, $P_g$ increases non-monotonically with $T_\text{LZ}$. This can be interpreted in terms of the competition between thermal excitation around the minimum gap and relaxation after the minimum gap at intermediate and long time scales, as seen in previous numerical studies of dissipative LZ ~\cite{nalbach_2009_landauzenertransitionsdissipative,nalbachAdiabaticMarkovianBathDynamics2014,arceci_2017_dissipativelandauzenerproblem}. As the total sweep time increases, $P_g$ first increases following the coherent limit, until the qubit has enough time to thermalize around the minimum gap and $P_g$ starts to decrease toward $0.5$, as $k_BT\gg\Delta$. Further increase of the total sweep time allows the qubit to relax after crossing the minimum gap, which leads to increasing $P_g$, as $k_BT\ll\epsilon^\t{final}\equiv 2I_p(\Phi_z^\t{final}-\Phi_z^\t{final})$. Since the instantaneous matrix element of the $\Phi_z$ noise is proportional to $\Delta/\sqrt{\Delta^2+\epsilon(t)^2}$, thermal relaxation after the minimum gap is slow, as $\epsilon^\t{final}\gg\Delta$, and is only significant at very long sweep times. 

Comparing the experimental data and AME simulation, we find qualitative agreement at large $\Phi_x$. The experimental data show smaller $P_g$ local maxima, which could indicate that the noise experienced by the qubit in the LZ sweep is larger than the extrapolated noise values based on qubit decoherence measurements. We note that at $\Phi_x=0.2$, the experimental data shows $P_g$ dropping slightly below 0.5 at intermediate sweep time, inconsistent with the interpretation that the system population at intermediate sweep time saturates toward the Boltzmann distribution near the minimum gap. This is likely due to imperfect state preparation at this particular $\Phi_x$ bias, caused by frequency collision with the readout resonator (see Supplementary Note 8). There is significant disagreement between the data and the AME simulation at smaller $\Phi_x$. Specifically, with regard to the $T_\text{LZ}$ dependence, while the AME predicts $P_g$ to nearly settle at 0.5, the experimental data shows $P_g$ to continue increasing with increasing sweep time. In fact, the experimental final ground state probability $P_g$'s for different $\Phi_x$ crosses at long times and the highest $P_g$ is obtained at the lowest $\Phi_x$. Furthermore, when examining the $\tau$ dependence, it can be seen that for $\tau\gg1$, where relaxation after the gap is expected to dominate, the simulated $P_g$ curves for different $\Phi_x$ collapse onto the same $\tau$ dependence. This is in contrast with the experimental data, where the $P_g$ curves shift left towards the coherent limit as $\Phi_x$ is reduced. These signatures indicate that the weak-coupling limit breaks down as the minimum gap $\Delta$ is reduced. 

To understand the data in the strong-coupling limit, we use the polaron-transformed master equation (PTRE)~\cite{chen_2022_hamiltonianopenquantum,smirnov_2018_theoryopenquantuma}. PTRE incorporates strong system-environment coupling by transforming into the dressed polaron frame and treats the tunneling parameter $\Delta$ perturbatively. The noise PSD in the polaron frame is separated into low- and high-frequency components. Particularly, the low-frequency part is characterized by two parameters,

\begin{align}
    W^2 &= 2I_p^2\int\frac{d\omega}{2\pi} S_{\Phi_z,\t{1/f}}^+(\omega),\,\text{and}\\
    \epsilon_p&=2I_p^2\int\frac{d\omega}{2\pi}\frac{S_{\Phi_z,\t{1/f}}^-(\omega)}{\hbar\omega},
\end{align}
which are known as the MRT width and reorganization energy respectively~\cite{amin_2008_macroscopicresonanttunneling}. The functions $S_{\Phi_z}^{-(+)}$ are the (anti-) symmetrized low frequency $\Phi_z$ noise. The PTRE is expected to hold when the environment-induced dephasing is much larger than the qubit's minimum energy gap, or $W\gg\Delta$. By integrating the $1/f$ noise obtained from dephasing time measurements, we obtain $W/h\approx 48-59\,\text{MHz}$. Assuming the environment is in thermal equilibrium, $\epsilon_p$ is related to $W$ via the fluctuation-dissipation theorem $\epsilon_p=W^2/2k_BT$. The result of PTRE simulations is shown in Figure~\ref{fig:ExpAMEPTRE}(e,f). It can be seen that for the nominal noise parameters found above, the final ground-state probability $P_g$ first increases with increasing sweep time, closely following the coherent LZ probabilities until around $P_g=0.5$ where it flattens. Through numerical experiments, it is also found that increasing $W$ while keeping $T$ unchanged, or equivalently increasing $\epsilon_p/W$, brings the PTRE results closer to the coherent LZ probabilities. In fact, it has been demonstrated previously that for a flux qubit strongly coupled to low-frequency noise, the LZ transition probability recovers the coherent LZ probability when $\epsilon_p\gg W$~\cite{johansson_2009_landauzenertransitionssuperconducting}.

In the experimental data, the transition probabilities approach the coherent limit as $\Phi_x$ and equivalently $\Delta$ is reduced, indicating a strong coupling between the qubit and the environment. At the smallest $\Phi_x$, the measured $P_g$ does not flatten near 0.5, contrasting the PTRE prediction with nominal noise values, but is closer to the PTRE prediction with larger MRT width $W$. This suggests that the noise seen by the qubit is larger than the integrated $1/f$ noise. This is not entirely surprising, given that previous MRT measurements on superconducting flux qubits also revealed larger $W$ than the integrated power of $1/f$ flux noise~\cite{harris_2010_experimentaldemonstrationrobust,quintana_2017_observationclassicalquantumcrossover}(see Supplementary Note 7). 

\subsection*{Spin bath model for the crossover regime}
\begin{figure*}
    \centering
    \includegraphics{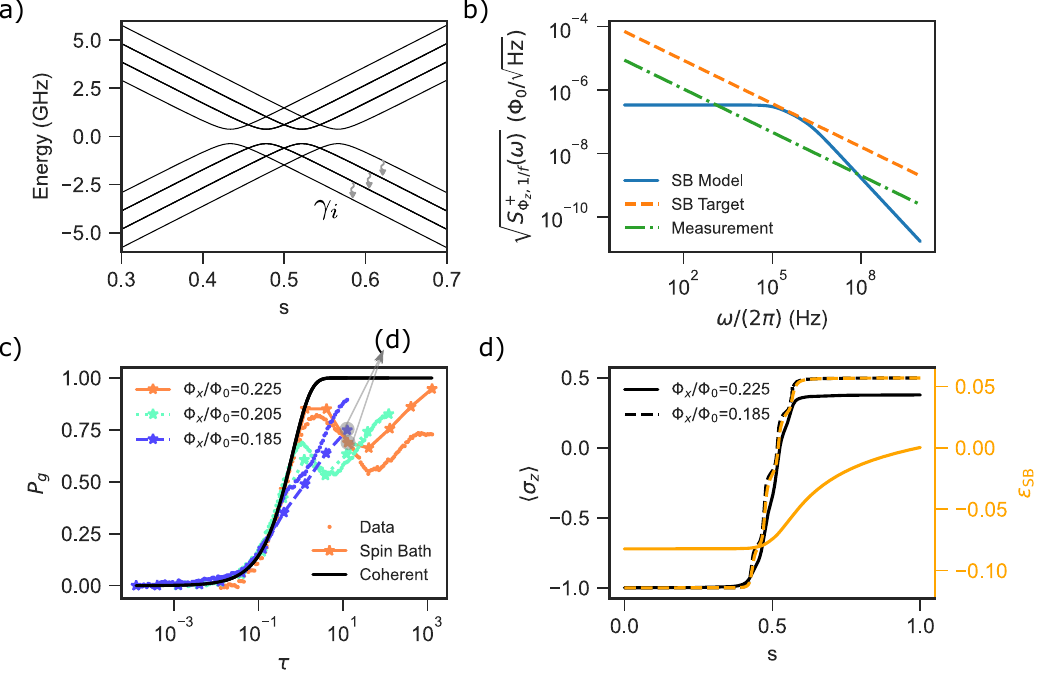}
    \caption{Simulation results of the spin bath model. (a) The energy spectra of the qubit going through an LZ transition, with 3 spins ferromagnetically coupled to the qubit, versus the normalized time $s\in[0, 1]$, for $\Phi_x/\Phi_0=0.225$. There are 4 distinct levels visible in the manifold of the qubit being in the left state, corresponding to 0, 1, 2, and 3 spins aligned with the qubit (with degeneracies of 1, 3, 3, and 1 respectively). Each of them anti-cross with the corresponding state in the manifold of the qubit being in the right state. Energy levels of states with different numbers of aligned spins cross each other, as there is no matrix element coupling them. The curly arrows indicate the paths for spin relaxation, allowing the spins to align with the qubit after tunneling. (b) The symmetrized $1/f^\alpha$ flux noise PSD as measured by the dephasing measurement (green dash-dotted line), the equivalent flux noise PSD generated by the spins (blue solid line), calculated by summing the individual Lorentzian contribution of each spin (see Supplementary Note 9), as well as the targeted noise PSD (orange dash line) that the spin bath is set to match, which has 8 times larger amplitude than the measured. (c)  The simulated final ground state probabilities versus dimensionless sweep time $\tau$, using the same spin bath parameters as in panels (a, b). (d) The evolution of the qubit polarization (black line) and effective bias by the spin bath (orange line), versus the normalized time $s$, for $\Phi_x=0.185$ (dashed line) and $\Phi_x=0.225$ (solid line) respectively. Both plots correspond to $\tau\approx 10$ (see text for discussion).}
    \label{fig:MultiSpinBath}
\end{figure*}

To further understand the result in the crossover regime, we propose a hybrid noise model, that incorporates both the weakly-coupled Markovian noise and the strong low-frequency noise through an explicit spin bath. The motivation for a spin bath is two-fold. First, spin impurities, characterized by logarithmically distributed relaxation times, are one of the main candidates for flux noise on superconducting circuits~\cite{koch_2007_modelfluxnoise,sendelbach_2008_magnetismsquidsmillikelvin,anton_2013_magneticfluxnoise,lanting_2020_probingenvironmentalspin,rower_2023_evolutionfluxnoise}. Second, polarization of the bath spins naturally lead to the concept of reorganization energy~\cite{lanting_2020_probingenvironmentalspin}, which is useful in connecting to the strong coupling PTRE model. To keep the model numerically feasible, we consider a spin bath made of $N_s=3$ spins coupled to the qubit via the interaction term
\begin{align}
    H_{qS} = \sum_i^{N_s}J_i\sigma_z\tau_{z,i},
\end{align}
where $\tau_{z,i}$ is the $i$'th bath spin's Pauli Z operator and $J_i$ is its coupling strength to the qubit. The bath spins do not have internal dynamics, but each of them is transversely coupled to its own environment, at a temperature that we assume to be the same as the qubit's environment, $T$. Each of these secondary environments leads to thermal transitions between the spin states, with depolarization rate $\gamma_i$. For appropriately chosen distributions of $\gamma_i$ and $J_i$, the noise PSD due to the spins effectively represents low-frequency noise with a specified noise amplitude in the chosen frequency range \cite{shnirman_2005_lowhighfrequencynoise}(see Supplementary Note 9). 

The spin bath model is simulated using the AME, with the high-frequency noise coupled to the qubit defined in the same way as the single-qubit AME. As a proof-of-concept demonstration, we choose a model with three spins, with uniform ferromagnetic coupling strength $J$, and $\gamma_i$ distributed in the range of $1-10~\t{MHz}$ to match the $1/f^\alpha$ noise power in this range. The range of $\gamma_i$ corresponds to an intermediate frequency range out of the full frequency span over which the $1/f^\alpha$ noise PSD is expected to hold, from sub-$\t{Hz}$ to about $1~\t{GHz}$ (see Supplementary Note 9). This choice is based on the understanding that noise processes slower than the adiabatic timescale do not affect the transition probability~\cite{kayanuma_1984_nonadiabatictransitionslevel}, and the dominant effect of fast noise is thermal transitions, which is accounted for by the environment directly coupled to the qubit, as is the case in the single-qubit AME. 

As shown in Fig.~\ref{fig:MultiSpinBath}, it is found that if the target $1/f$ noise amplitude is about 8 times larger than the noise amplitude deduced from decoherence measurements (corresponding to $J_i/h=J/h=-0.09~\text{GHz}$), the simulated ground state probabilities versus sweep time at different $\Phi_x$ qualitatively matches the experimental results. 
Specifically, for large $\Phi_x$ or $\Delta$, the spin bath simulation results display the non-monotonic dependence of ground state probability $P_g$ versus the dimensionless sweep time $\tau$, and is almost indistinguishable from the single qubit AME result. However, at smaller $\Phi_x$ or $\Delta$, the spin bath simulation result displays a monotonic increase of $P_g$ with increasing sweep time $\tau$, which is similar to the PTRE. The spin bath simulation results also differ from both AME and PTRE in that the $P_g$ curves for different $\Phi_x$ at large $\tau$ do not collapse.

Further intuition about the spin bath model can be obtained by observing the evolution of the polarization of the qubit and the spin bath during the LZ sweep. The collective effect of the spins can be captured by the parameter
\begin{align}
    \epsilon_\text{SB}=-\sum J_i\langle\tau_{z,i}\rangle,
\end{align}
which is the effective longitudinal bias applied by the spins on the qubit, analogous to the reorganization energy $\epsilon_p$ in the PTRE model (see Supplementary Note 9). The instantaneous change of $\epsilon_\text{SB}$ during the sweep is to be compared with the change in qubit polarization $\langle\sigma_z\rangle$, as shown in Fig.~\ref{fig:MultiSpinBath}(d) for two different values of $X$-flux biases in the adiabatic limit. We can further denote the temporal width of the change of the qubit polarization as the tunneling time $T_t$, which in the adiabatic limit is approximately $\Delta/v$~\cite{mullen_1989_timezenertunneling}. The tunneling time for the parameter $\Phi_x=0.225(0.185), T_\t{LZ}=10^3(10^5)~\t{ns}$ is $T_t=108 (995)~\t{ns}$. It can be seen that for $\Phi_x=0.225$, the effective bias $\epsilon_\text{SB}$ changes slowly after the qubit has tunneled. Therefore the spin bath presents negligible influence on the qubit dynamics. For $\Phi_x=0.185$, the change in $\epsilon_\text{SB}$ almost overlaps with the changes in qubit polarization $\langle \sigma_z\rangle$. This is because the spin depolarization rates become shorter as compared to the tunneling time of the qubit as $\Phi_x$ or $\Delta$ reduces. In other words, for small $\Delta$, the spins quickly reorganize themselves to align with the new qubit polarization as the qubit tunnels. This fast change in $\epsilon_\text{SB}$, together with its relatively large magnitude as compared to $\Delta$, effectively shifts the qubit away from the anti-crossing as soon as the qubit has tunneled to the opposite polarization. Away from the anti-crossing, the qubit rarely experiences thermal excitation, hence the non-monotonic $P_g$ dependence disappears and $P_g$ only increases with sweep time as it has more time to complete the tunneling. If however the noise is not strong and does not induce a large enough $\epsilon_\text{SB}$ to shift the qubit away from the anti-crossing, thermal excitation would still occur, and $P_g$ would barely increase above 0.5, similar to the single-qubit AME when $\Delta$ is small (see Supplementary Note 9). 

\section*{Discussion}\label{sec:LZDiscussion}
The dissipative LZ transition studied here can be considered a toy model for dynamics in quantum annealing, where small gap anti-crossings between the lowest two energy eigenstates are expected to play an important role~\cite{amin_2008_thermallyassistedadiabatic,boixo_2016_computationalmultiqubittunnelling}. Our results thus contribute to the understanding of the role of open-system effects in a quantum annealer~\cite{dickson_2013_thermallyassistedquantum,albashDecoherenceAdiabaticQuantum2015,boixo_2016_computationalmultiqubittunnelling,mishra_2018_finitetemperaturequantum,bando_2022_breakdownweakcouplinglimit,king_2022_coherentquantumannealinga}. In the weak-coupling limit, thermal relaxation could help the system to reach the ground state. However, for hard problems, the minimum gap is expected to close, which freezes thermal relaxation after the system has passed the minimum gap~\cite{albash_2012_quantumadiabaticmarkovian,amin_2015_searchingquantumspeedup,boixo_2016_computationalmultiqubittunnelling}. On the other hand, the experiment and the PTRE model both show that higher ground state probabilities can be achieved in the intermediate to strong coupling limit, as compared to the weak-coupling limit, when $\epsilon_p/W$ is large enough. However, this is likely specific to the LZ tunneling problem, where the final (right) well becomes much lower than the initial (left) well, allowing tunneling to occur despite the large environment bias preferring the initial well. In the context of hard annealing problems with many quasi-degenerate energy levels, tunneling dynamics of the system could be suppressed due to the environment biasing the energy levels in the strong coupling limit. 

The spin bath simulation suggests that only noise that is fast enough as compared to the qubit dynamics will contribute to the strong coupling effect. It is thus expected that in fast annealing, slow noise will not affect the annealing performance, as there is not enough time for a state of definite polarization to be formed. This is in line with the recent demonstration of coherence in fast annealing of flux-qubit-based quantum annealers, on the order of $10~\t{ns}$, where the environment is too slow to affect the distribution of final states~\cite{king_2022_coherentquantumannealinga,king_2023_quantumcriticaldynamicsa}. This immunity to slow noise also points to an important advantage of analog quantum processors over gate-based quantum processors, where in the latter slow noise always causes dephasing and needs to be actively corrected for. 

An interesting future direction would be environment engineering techniques to shift the noise at intermediate frequencies to low frequencies, which has been recently demonstrated by applying a magnetic field in-plane to the superconducting circuit~\cite{rower_2023_evolutionfluxnoise}. Further development of such techniques likely requires an improved understanding of the physical origins of flux noise. It should be noted that in the simulation, the noise power required to obtain agreement with the experiment is larger than estimated from dephasing time measurements. This could be due to the sensitivity of the measurement to different noise frequencies, and the non-Gaussian nature of the low-frequency noise. Future experiments could use different control protocols, such as repeated LZ crossing, and locally adiabatic control protocols, which likely provide sensitivity to noise at different frequencies. These experiments, in combination with noise spectroscopy techniques~\cite{bylander_2011_noisespectroscopydynamical}, could be used to further settle the actual noise power seen by the qubit, and elucidate the physical origins of low-frequency flux noise.  

In summary, we experimentally characterized the LZ transition probability in a superconducting flux qubit with a wide range of sweep velocities $v$ and minimum gap sizes $\Delta$, and we showed a crossover from weak to strong coupling to flux noise. We found that for large gap $\Delta$, the competition between adiabaticity and environment-induced thermalization leads to non-monotonic dependence of the final ground state probability $P_g$ on sweep time, which can be reproduced by a weak-coupling model, the AME. However, as $\Delta$ becomes smaller, the non-monotonicity gradually disappears and $P_g$ becomes closer to the coherent LZ transition rate, which is consistent with a strong-coupling model, the PTRE. We also explored a spin bath model that qualitatively reproduces the full range of experimental data. The spin bath model explicates that the crossover depends on the relative timescale between the tunneling time of the qubit, and the time taken for the environment to reorganize itself after the qubit has tunneled. Our work brings insights into the role of low-frequency noises in quantum tunneling, which is relevant to quantum annealing and more broadly tunneling phenomena in quantum chemistry and biology. Future extensions of this work include direct measurement of the environment reorganization energy and timescale in different control protocols and the investigation of the crossover region in multi-qubit settings.

\section*{Methods}
\subsection*{Device fabrication}
The device is fabricated at MIT Lincoln Laboratory using a multi-tier fabrication process, which consists of a high-coherence qubit tier, an interposer, and a multi-layer control tier~\cite{rosenberg_2017_3dintegratedsuperconducting,yost_2020_solidstatequbitsintegrated}. Our device only uses the qubit and interposer tiers. The qubit tier contains high-quality Al forming the qubit loop and capacitance, as well as $\text{Al}/\text{Al}\text{O}_\text{x}/\text{Al}$ junctions. The interposer tier contains the tunable resonator and the control lines. They are bump-bonded together in a flip-chip configuration using indium bumps.

\subsection*{Qubit control and readout}
The qubit $x, z$ flux biases are each controlled using on-chip flux bias lines, with current supplied by a DC voltage source and a fast AWG. The DC source has a stronger coupling to the qubit loops to achieve flux biasing of more than a flux quantum. The AWG has weaker coupling to the flux biases, corresponding to a bias range of about $100~\text{m}\Phi_0$. 

Before the LZ sweep, the qubit needs to be prepared in its ground state. Passive cooling is not possible when the tunneling barrier between persistent current states is large. We use a cooling protocol similar to that demonstrated in Ref.~\cite{valenzuela_2006_microwaveinducedcoolingsuperconducting}. Using the large tunneling amplitude between the ground state of the higher well and the excited state of the opposite well, residual excited populations can be adiabatically transferred to the lower well and the system quickly relaxes to the ground state. Additional details of the energy spectrum and this cooling method is discussed in Supplementary Note 6. 

The qubit state is read by measuring the state-dependent transmission through the tunable resonator. When doing spectroscopy and coherence measurements, the resonator is biased at a flux-insensitive position ($0$ flux in the SQUID), and readout is in the qubit energy eigenbasis, due to dispersive interaction between the qubit and resonator. During the LZ measurement, the resonator is biased at a flux-sensitive position ($-0.15~\Phi_0$ flux in the SQUID), which allows measuring the qubit's persistent current states.  Since the readout point of the LZ experiment is far from the qubit's symmetry point, the persistent current basis and energy eigenbasis of the qubit nearly coincide, with more than 99\% overlap, and we do not distinguish the two bases at the readout point.   

\subsection*{Noise characterization}
The qubit is capacitively coupled to a microwave line. This allows spectroscopy, qubit relaxation (T1) and qubit dephasing (T2) measurement when $\Delta/h\gtrsim 1~\text{GHz}$. The flux-dependent qubit transition frequencies are used to fit a complete circuit model of the device, which is used to compute $I_p$ and $\Delta$ at different $\Phi_x$~\cite{kerman_2020_efficientnumericalsimulation}. The circuit model is also used to compute the noise sensitivity of the qubit at different flux biases. Combining the noise sensitivity with the T1, T2 measurement results allows us to determine the noise PSD, which is used in the master equation simulation for LZ transition. More details of the noise characterization is discussed in a separate pubcalication~\cite{trappen_2023_decoherencetunablecapacitively},The noise parameters used are given in Supplementary Note 7.

\subsection*{Master equation simulation}
The master equation simulations were performed using HOQST, a Julia package for open-system dynamics with time-dependent Hamiltonians~\cite{chen_2022_hamiltonianopenquantum}. The simulation takes the qubit Hamiltonian in Eq.~(\ref{eq:QubitHam}), with $I_p, \Delta$ given by the circuit model. The qubit-bath interaction considered is  
\begin{align}
    H_{qb}=-I_p\sigma_z\otimes Q_{\Phi_z},
\end{align}
where $\sigma_z$ acts on the qubit and $Q$ acts on the environment degrees of freedom which causes the qubit $z$ loop flux noise.

The AME is a time-dependent version of the frequency-form Lindblad equation. Although a rigorous upper bound on the error of the AME is only small in the adiabatic limit~\cite{mozgunov_2020_completelypositivemaster}, it is likely that the error is still small in the non-adiabatic limit, as shown in previous work where the master equation results are compared to numerically exact path integral based simulation results~\cite{nalbachAdiabaticMarkovianBathDynamics2014,arceci_2017_dissipativelandauzenerproblem}. Specifically, the form used in this work is the one-sided AME that first appeared in~\cite{albash_2012_quantumadiabaticmarkovian}, given by
\begin{align}
    &\dot\rho_q(t) = -\frac{i}{\hbar}[H_q(t), \rho_q(t)]\nonumber\\
    &+\frac{1}{\hbar^2}\sum_{\omega}\Gamma_{\Phi_z}(\omega)\nonumber\left[L_{\omega}(t)\rho_q ,I_p\sigma_z \right]+h.c.\label{eq:AME},
\end{align}
where $\rho_q$ is the reduced density matrix of the qubit,
\begin{align}
    \Gamma_{\Phi_z}(\omega)&=\frac{1}{2}S_{\Phi_z}(\omega)+i\gamma_{\Phi_z}(\omega)\\
    \gamma_{\Phi_z}(\omega)&=\frac{1}{2\pi}\int_{-\infty}^\infty S_{\Phi_z}(\omega^\prime)\mathcal{P}\left(\frac{1}{\omega-\omega^\prime}\right)d\omega^\prime,
\end{align}
with $\mathcal{P}$ denoting the Cauchy principal value. The operators $L_\omega$ are given by
\begin{align}
L_\omega(t)=\sum_{E_\beta-E_\alpha=\omega}\bra{\alpha(t)}I_p\sigma_z\ket{\beta(t)}\ket{\alpha(t)}\bra{\beta(t)},
\end{align}
where $E_{\alpha(\beta)}$ and $\ket{\alpha(\beta)}$ are the system's instantaneous energy eigenvalues and eigenstates, and $\alpha,\beta\in\{g,e\}$. 

The PTRE is a model that accounts for strong system-environment coupling and has been found to explain experimental data in quantum annealers coupled strongly to low-frequency noise~\cite{johansson_2009_landauzenertransitionssuperconducting, smirnov_2018_theoryopenquantuma}. The form of PTRE we use here has the Lindblad form, given by
\begin{align}
    &\dot{\tilde{\rho}}_q(t) = -\frac{i}{\hbar}[\tilde{H}_q(t), \tilde{\rho}_q(t)]\nonumber\\
    &+\frac{1}{\hbar^2}\sum_{\omega,\lambda}\tilde{S}(\omega)\nonumber\left[\tilde{L}_{\omega,\lambda}\tilde{\rho}_q \tilde{L}_{\omega,\lambda}^\dagger-\frac{1}{2}\left\{\tilde{L}_{\omega,\lambda}^\dagger\tilde{L}_{\omega,\lambda},\tilde{\rho}_q\right\}\right]\label{eq:PTRE},
\end{align}
where tilde is used to denote operators in the polaron frame. Specifically,
\begin{align}
    \tilde{H}_q &= -I_p (\Phi_z-\Phi_z^\text{sym})\sigma_z,\\
    \tilde{L}_{\omega,\lambda\in\{+,-\}} &= \frac{\Delta}{2}\sum_{E_i-E_j=\hbar\omega}\bra{i}\sigma_{\lambda}\ket{j}\ket{i}\bra{j},
\end{align}
where $\sigma_{+(-)}$ are the qubit Pauli raising and lowering operators and $i,j\in\{0,1\}$ are the qubit persistent current state index. The low- and high-frequency parts of the noise give a convolutional form for the PSD in the polaron frame

\begin{align}
    \tilde{S}(\omega)&=\hbar^2\int\frac{\text{d}\omega^\prime}{2\pi}G_L(\omega-\omega^\prime)G_H(\omega^\prime),
\end{align}
where $G_L$ and $G_H$ are contributions from the low- and high-frequency noise respectively, given by
\begin{align}
    G_L(\omega)&=\sqrt{\frac{\pi}{2\hbar^2W^2}}\exp\left[-\frac{(\hbar\omega-4\epsilon_p)^2}{8\hbar^2W^2}\right]\\
    G_H(\omega)&=\frac{4S_{\Phi_z,\text{ohmic}}(\omega)I_p^2}{\hbar^2\omega^2+4S_{\Phi_z,\text{ohmic}}(0)I_p^2}.
\end{align}

\subsection*{Spin bath simulation}
In the spin bath model, the Hamiltonian of the system qubit, the spin bath, together with their respective environment is given as
\begin{align}
    H &= H_q+ H_S+ H_{qS} + H_{qb} + H_{SB} + H_b + H_{B},\\
\end{align}
with
\begin{align}
    H_{qS} &= \sum_i^{N_s}J_i\sigma_z\tau_{z,i},\\
    H_S &= 0,\\
    H_{SB} &= \sum_i^{N_s}\tau_{x,i} Q_i^\prime,~\text{and}\\
    S_{Q_i^\prime}(\omega)&=\hbar^2\lambda_i\frac{1}{1+\exp{(-\beta\hbar\omega)}}\exp{\left(-\frac{\omega}{\omega_c}\right)}.
\end{align}
Here $H_{b}$ is the qubit environment and $H_{B}$ is the secondary environment coupled to the spins. The operator $Q_i^\prime$ is an operator of the $i^\prime$th secondary environment that is coupled to spin $i$. The noise PSD of the $Q_i^\prime$ operator is nearly white noise, with strength $\lambda_i$, inverse temperature $\beta$ and cutoff frequency $\omega_c$ assumed to be the same as the qubit high-frequency environment. This can be thought of as a quantum extension of simulating classical $1/f$ noise with two-level fluctuators~\cite{dutta_1981_lowfrequencyfluctuationssolids}, The crucial difference here that distinguishes these spins from classical two-level fluctuators is that the relaxation and excitation rates obey detailed balance. Indeed, if the spins are sufficiently weakly coupled to the qubit and the energy difference between the spin up and down states are zero (or if the spin environment has infinite temperature), the spin has the same relaxation and excitation rate, $\lambda/2$, acting exactly like a classical two-level fluctuator.

The spin bath parameters $J_i$ and $\lambda_i$ are chosen based on the measured $1/f$ flux noise strength. Specifically, the noise strength $\lambda_i$ determines the spin depolarization rate and is chosen based on the range of noise frequency of interest. To determine the coupling strength $J_i$, we consider the effective noise PSD contributed by each spin~\cite{shnirman_2005_lowhighfrequencynoise}, given by
\begin{align}
    S_i(\omega)&=(1 -\langle\tau_i\rangle)\frac{2\gamma_i J_i^2}{\omega^2+\gamma_i^2},
\end{align}
where $\langle\tau_i\rangle$ is the expectation value of spin $i$'s longitudinal polarization. The coupling strength $J_i=J$ is determined by summing the PSDs of individual spins and comparing with the target noise amplitude. We note that the spin's effective PSD is strictly only valid when it is weakly coupled to the qubit. However, it serves as a useful guide in choosing the spin bath parameters. Further simulation of the spin bath model and its comparison with the MRT parameters are given in the Supplementary Note 9.

\section*{Data availability}
The data generated in this study have been deposited in the Figshare database, which can be accessed via the link \url{https://doi.org/10.6084/m9.figshare.26531362}~\cite{figshare}. 

\section*{Code availability}
Analysis and simulation codes that support the findings of this study are available from the corresponding authors upon request. 
%REFERENCES

\begin{acknowledgments}
We thank the members of the Quantum Enhanced Optimization (QEO)/Quantum Annealing Feasibility Study (QAFS) collaboration for various contributions that impacted this research. In particular, we thank D. Ferguson for fruitful discussions of experiments, A. J. Kerman for guidance on circuit simulations and design, and R. Yang for related work on circuit modeling and useful discussions. We also thank fruitful discussions with W. Strunz, V. Link, F. S. Kahlert and D. Segal on the open quantum system models. We gratefully acknowledge the MIT Lincoln Laboratory design, fabrication, packaging, and testing personnel for valuable technical assistance. The theoretical modeling of the work benefited from the high-performance computing cluster provided by SHARCNET (sharcnet.ca) and the Digital Research Alliance of Canada (alliancecan.ca). The research is based upon work supported by the Office of the Director of National Intelligence (ODNI), Intelligence Advanced Research Projects Activity (IARPA) and the Defense Advanced Research Projects Agency (DARPA), via the U.S. Army Research Office contract W911NF-17-C-0050. The views and conclusions contained herein are those of the authors and should not be interpreted as necessarily representing the official policies or endorsements, either expressed or implied, of the ODNI, IARPA, DARPA, or the U.S. Government. The U.S. Government is authorized to reproduce and distribute reprints for Governmental purposes notwithstanding any copyright annotation thereon.

\end{acknowledgments}
\section*{Author Contributions Statement} 
R.T. performed the experiments. R.T. and X.D. performed the data analysis. X.D. performed the numerical simulations. H.C. provided guidance on the numerical simulations and theoretical models. D.M., M.A.Y., A.J. Martinez, and Y.T. designed the device. S.N. provided feedback on the device design. E. M. explored alternative explanations of the experimental data within the AME model. J.G. J.A.G. and X.D. performed earlier versions of the experiments on a different device and setup. D.M.T., J.A.G., S.D., and J.B. contributed to the development of the experiment infrastructure. R.D., A.J.Melville, B.M.N., and J.L.Y. developed the fabrication process and fabricated the device. C.H., K.S., and S.J.W. contributed to the fridge and electronics operation. W.O. supervised the QEO/QAFS effort from Lincoln lab, K.M.Z. led the coordination of the QEO/QAFS experimental effort, D.L. led the QEO/QAFS program and A.L. proposed and supervised this work. All authors were involved in the discussion of experiments and data analysis. X.D., R.T., and A.L. wrote the paper with feedback from all authors. 

\section*{Competing Interests Statement}
The authors declare no competing interests.

%When using a bibtex file (the only option you should consider at the draft stage), use the section below.
%\bibliography{main}%bibtex file, use path relative to the folder
%\bibliographystyle{naturemag}%
% this is normally done when choosing journal in document class
%styles to use: 
%for APS 
%apsrev4-1.bst(APS journals using a numeric citation style, i.e., all but RMP)
%apsrmp4-1.bst (author/year style citations for RMP)
%aipauth4-1.bst (AIP journal using an author/year citation style)
%aipnum4-1.bst (AIP journals using a numeric citation style)
%for Nature journals: naturemag

%the section below is used when references are inserted in the file. DO NOT USE this option unless needed. Note many (most?) journals require that the bibliography is contained in the file itself. In this case, work with bibtex until the submission phase. Once ready, comment the lines above, that is \bibliography{} and \bibliographystyle{}. Then check the bbl file (generated by latex). It has all the compiled bibliography entries, formatted according to the chosen style. The entire file content has to be copied and it has to replace the three comment lines below (this code section has to start with \begin{thebibliography} and end with \end{thebibliography})
%\begin{thebibliography}{20}
%biblio text here
%\end{thebibliography}

%%%%%%%%%% Merge with supplemental materials %%%%%%%%%%

\end{document}

%\widetext
%\linenumbers
%\pagebreak
%\newpage
%\begin{center}
%\bf{\large Supplementary Information: Dissipative Landau-Zener tunneling: crossover from weak to strong environment coupling}
\title{Supplementary Information}
\maketitle

%%%%%%%%%% Merge with supplemental materials %%%%%%%%%%
%%%%%%%%%% Prefix a "S" to all equations, figures, tables and reset the counter %%%%%%%%%%
\setcounter{equation}{0}
\setcounter{figure}{0}
\setcounter{table}{0}
\setcounter{page}{1}
%\makeatletter
\renewcommand{\theequation}{S\arabic{equation}}
\renewcommand{\thesection}{Supplementary Note \arabic{section}}
\renewcommand{\figurename}{Supplementary Fig.}
\renewcommand{\tablename}{Supplementary Tab.}
\renewcommand{\bibnumfmt}[1]{[S#1]}
\renewcommand{\citenumfont}[1]{S#1}
\tableofcontents

\section{Wiring}\label{S_sec:Wiring}

\begin{figure}[b]
    \centering
    \includegraphics{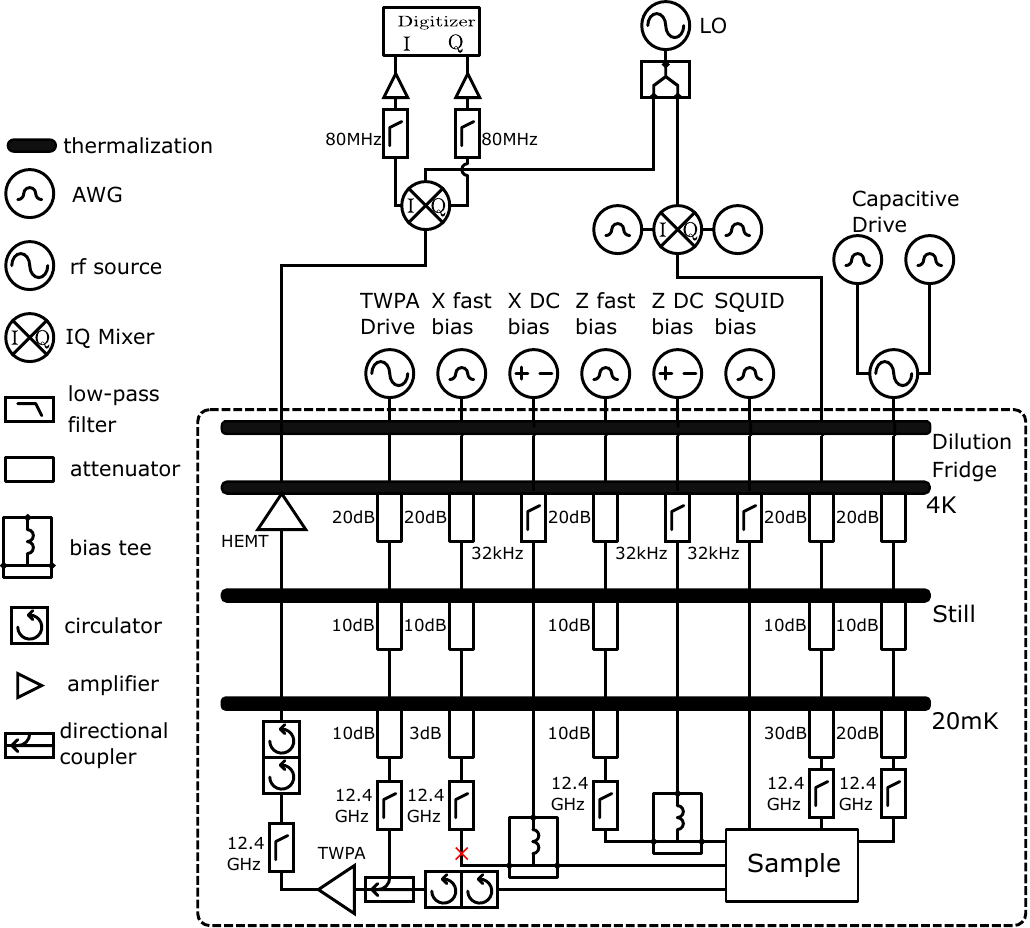}
    \caption{Schematic of the setup used for experiments. The open fast X-bias line is marked in red cross.}
    \label{S_fig:FridgeWiring}
\end{figure}

In Supplementary Figure~\ref{S_fig:FridgeWiring} we present a diagram of the room-temperature and cryogenic setup used in the experiments. The qubit flux biases for the $x$ and $z$ loops, $\Phi_x$ and $\Phi_z$, are controlled by currents generated by a DC voltage source and a high bandwidth source, combined through a bias tee at the mixing chamber stage of the dilution refrigerator. The DC biases are supplied by Yokogawa GS 200 voltage sources through twisted pair cables and low-pass filtered to below 32~kHz to minimize noise. The high bandwidth components of the current are supplied by a 1~GHz arbitrary waveform generators (AWGs) (Keysight M3202A), through coaxial cables. The fast line coupling to $\Phi_x$ suffered from a cold open during the cooldown and is not used for the Landau-Zener experiment. The attenuators and filters on the cable are chosen to allow a sufficiently large current range for driving Rabi oscillation and performing annealing experiment, while minimizing decoherence due to thermal and electronic noise. The resonator SQUID bias is supplied by an AWG (Keysight M3202A), but through twisted pair cable. The qubit can be driven with a capacitively coupled driving line,  supplied by an rf source with an integrated IQ modulator, through a coaxial cable. The readout signal is amplified by a travelling wave parametric amplifier (TWPA)~\cite{macklin_2015_quantumlimitedjosephsontravelingwave} at the mixing chamber stage of the fridge, followed by a high-electron mobility transistor amplifier (HEMT) at the 4K stage, and room temperature amplifiers, before being processed by a field-programmable gate array (FPGA) digitizer (Keysight M3102A).

\section{Crosstalk Calibration}\label{S_sec:Crosstalk}
DC flux crosstalk between different bias lines and flux loops are calibrated using the CISCIQi method developed in Ref.~\cite{dai_2021_calibrationfluxcrosstalk}. We first measure the flux bias dependent resonator spectrum and the crosstalk into the resonator from other bias lines. This allows us to fix the resonator bias to measure the qubit-bias dependent transmission through the resonator. The procedure is iterated a few times until the crosstalk is compensated to within $1\,\text{m}\Phi_0$ accuracy. The full crosstalk matrix is shown in Supplementary Figure~\ref{S_fig:QubitCircuit}(a).

Crosstalk from the fast pulses sent to the qubit $z$ loop to other loops are not compensated due to bandwidth limitations on other bias lines. For the small pulse amplitude used for the Landau-Zener sweep, reaching up to $10\,\text{m}\Phi_0$, the induced flux on the $x, r$ loop should be inconsequential.

\section{Circuit Model}\label{S_sec:CircuitModel}
The device used in this experiment consists of a capacitively-shunted flux qubit coupled to a tunable rf-SQUID terminated resonator. The coupling between these two circuits is done via the mutual inductance between the qubit $z$ loop and the rf-SQUID loop. A lumped element representation of the qubit and resonator circuit is shown in Supplementary Figure~\ref{S_fig:QubitCircuit}(b). The qubit eigenstates and eigenvalues are obtained using the numerical tools developed in Ref.~\cite{kerman_2020_efficientnumericalsimulation}. The circuit persistent current $I_p$ and gap $\Delta$ were verified experimentally via qubit spectroscopy, as shown in Supplementary Figure~\ref{S_fig:QubitCircuit}(c). The simulated $I_p$ and $\Delta$ values are plotted as a function of $\Phi_x$ in Supplementary Figure~\ref{S_fig:QubitCircuit}(d).

The parameters of the resonator rf-SQUID were determined from fitting the experimental values of the resonator frequency versus resonator flux bias $\Phi_r$, shown in Supplementary Figure~\ref{S_fig:Resonator}(a). The resonator model allows for extraction of the value of the screening current in the rf-SQUID, which was used to determine the shift in the symmetry point of the qubit, induced by the rf-SQUID when biased away from zero. In order to confirm this value experimentally, spectroscopy curves were taken at a fixed value of $\Phi_x$ at two different values of resonator bias $\Phi_r$ at zero and -0.15, the later being the value used for persistent current readout. The two spectroscopy curves are shown in Supplementary Figure~\ref{S_fig:Resonator}(b). The circuit parameters are summarized in Supplementary Table~\ref{S_tab:CircuitParams}.
\begin{table}
    \centering
    \begin{tabular}{|c|c|}
    \hline
         $L_r$ & 16.4pH \\
        \hline
         $L_l$ & 16.4pH \\
         \hline
         $L_z$ & 799.8pH \\
         \hline
         $I_\text{cr}$ & 77.6nA \\
         \hline
         $I_\text{cl}$ & 89.1nA \\
         \hline
         $I_\text{c1}$ & 244.7nA \\
         \hline
         $I_\text{c2}$ & 244.7nA \\
         \hline
         $C_\text{cd}$ & 0.4fF\\
         \hline
         $C_\text{ad}$ & 3.6fF\\
         \hline
         $C_\text{ac}$ & 0.04fF\\
         \hline
         $C_\text{ab}$ & 14.2fF\\
         \hline
         $C_\text{gd}$ & 44.1fF\\
         \hline
         $C_\text{gc}$ & 0.2fF\\
         \hline
         $C_\text{ga}$ & 141.4fF\\
         \hline
         $C_\text{gb}$ & 90.0fF\\
         \hline
         SQUID inductance $L_s$ & 238.7pH\\
         \hline
         SQUID junction $I_s$ & 1187nA\\
         \hline
         resonator length $l_r$ & 3.364mm\\
         \hline
         Mutual between qubit and SQUID $M$ & 60.3pH\\
    \hline
    \end{tabular}
    \caption{Fit model best parameters for the qubit and resonator circuits.}
    \label{S_tab:CircuitParams}
\end{table}
\begin{figure}
    \centering
    \includegraphics{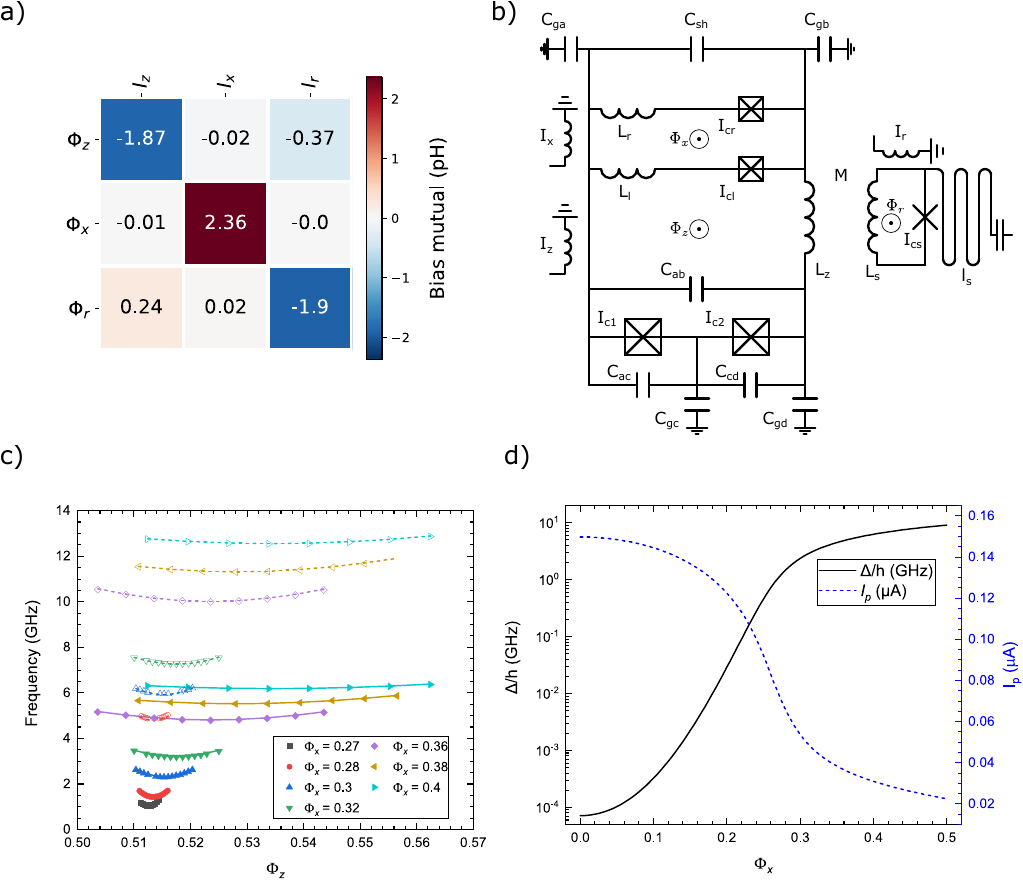}
    \caption{a) Matrix of mutual inductances between on chip bias lines and flux loops.. b) Circuit diagram for the qubit and coupler. c) Experimentally measured qubit transition frequency and simulated qubit transition frequencies for the best fit parameters, as a function of the biases $\Phi_x$ and $\Phi_z$, and comparison with experiment. Filled symbols (solid lines) correspond to the experimentally obtained (simulated) transition frequencies between the ground and first excited state. Open symbols (dashed lines) correspond to the experimental (simulated) transition frequencies between the ground and second excited state. d) Simulated minimum gap (left axis, solid curve) and persistent current (right axis, dashed curve) values as a function of the bias $\Phi_x$.
}
    \label{S_fig:QubitCircuit}
\end{figure}

\begin{figure}
    \centering
    \includegraphics{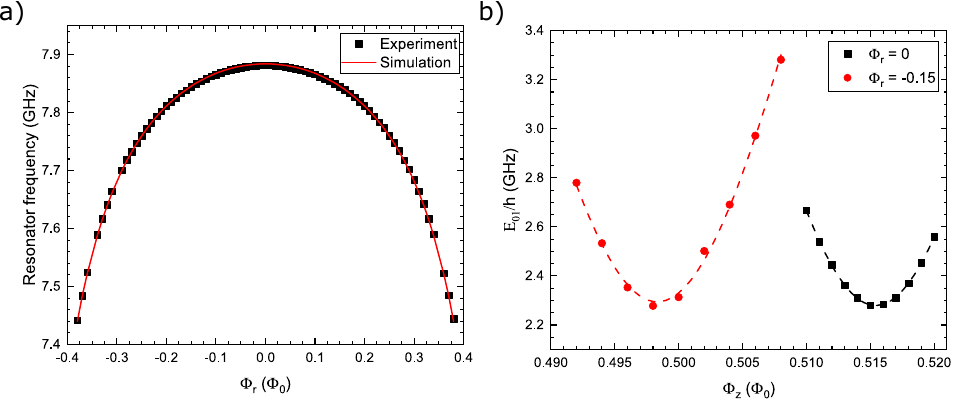}
    \caption{a) Readout resonator resonance frequency vs bias: experiment (solid points) and simulation with best fit parameters (solid line). b) Experimentally measured shift of the qubit z symmetry point due to the SQUID screening current. Solid symbols correspond to qubit frequencies determined from spectroscopy, and dashed lines correspond to a fit using a two-level system anticrossing relation.
}
    \label{S_fig:Resonator}
\end{figure}
\section{Fast flux line coupling characterization}\label{S_sec:FastFluxCoupling}
As described in Supplementary Note~\ref{S_sec:Wiring}, fast voltage pulses are applied to the qubit $z$ loop via a bias tee in order to control the flux bias of the qubit during the Landau-Zener sweep. In order to determine the transfer function between the voltage of the AWG and the flux fed to the qubit loop, a procedure using Ramsey interferometry was used.

The protocol is shown in Supplementary Figure~\ref{S_fig:FastFluxCoupling}(a). Two $\pi$/2 pulses are applied using the capacitively coupled waveguide, separated by the time $\tau_{\text{delay}}$. During the interval between the two $\pi$/2 pulses, a trapezoidal flux pulse with 1~ns rise and fall time is applied to the $z$ loop of the qubit. This flux pulse adiabatically changes the qubit frequency, depending on the pulse amplitude and duration, which induces an additional phase for the superposition created by the first $\pi/2$ pulse, inducing an oscillation whose period depends on the flux amplitude. An example is shown in Supplementary Figure~\ref{S_fig:FastFluxCoupling}(b) for the case with no flux pulse applied (i.e. pure Ramsey with no detuning) and with a 60 mV pulse applied. 

The above sequence is repeated at a range of pulse amplitudes such that the detuning varied between approximately from 0 to 60 MHz. This detuning versus flux pulse amplitude is shown in Supplementary Figure~\ref{S_fig:FastFluxCoupling}(c). Combining the detuning and the drive frequency gives the effective qubit frequency during the delay time in the Ramsey sequence. Comparing the qubit frequency versus flux pulse amplitudes allows us to deduce the voltage to flux conversion for the fast flux line.

\begin{figure}
    \centering
    \includegraphics{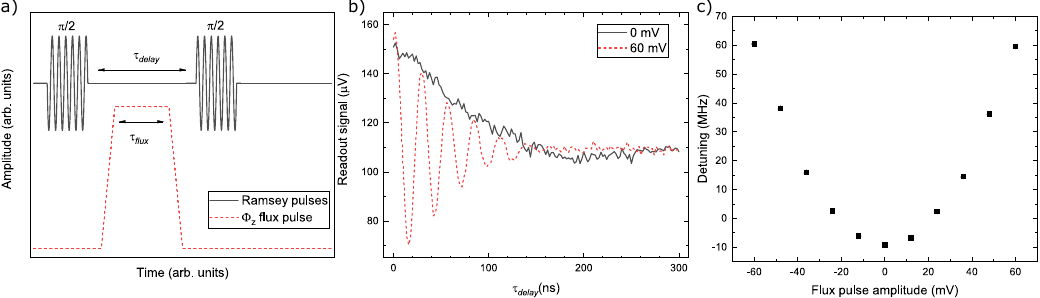}
    \caption{Bias line characterization using a Ramsey protocol. a) Pulse sequence for Ramsey measurements. The qubit is initialized in a superposition using a $\pi/2$ pulse, then a flux pulse applied to the qubit Z bias adiabatically detunes the qubit away from the symmetry point. The spacing between the two $\pi/2$ pulses $\tau_{\text{delay}}$ equal to the pulse duration $\tau_{\text{flux}}$ plus the rise and fall times. The qubit acquires a phase dependent on the amplitude and duration of the pulse. b) Ramsey oscillation curves with $V_z$ = 0 and 60 mV. c) Fitted detuning as a function of pulse amplitude. 
}
    \label{S_fig:FastFluxCoupling}
\end{figure}

\section{Fast flux line pulse distortion characterization}\label{S_sec:PulseDistortion}
During the Landau-Zener measurement sequence, time-dependent flux pulses are applied to the qubit via a bias tee, as discussed in ~\ref{S_sec:Wiring} and \ref{S_sec:FastFluxCoupling}. As the experiment involved the application of long pulses (duration  $>$ 1 $\mu$s), we characterize the transmission of the AC port of the bias tee to check for possible frequency dependent attenuation effects that would distort the pulse shape.

In order to check for distortion effects, we used an experimental protocol based on a Ramsey sequence, shown in Supplementary Figure~\ref{S_fig:FastFluxPulse}(a). Two microwave $\pi$/2 pulses are applied with a fixed delay time $\tau_\text{delay}$. The delay time is chosen to correspond to the inflection point of one of the Ramsey oscillations, so that the readout signal is maximally sensitive, and responds linearly to changes in qubit frequency and hence the flux pulse amplitude. The spacing between the Ramsey sequence and readout is fixed. In addition, a square pulse from $\Phi_z$ fast line is applied with duration $\tau_{\text{flux}}$. The position of the rising edge of the square pulse is varied, spanning a range of times relative to the Ramsey pulses, from the rising edge following the first $\pi/2$ pulse to preceding the first $\pi/2$ pulse. This is done by increasing $\tau_{\text{flux}}$ while keeping the falling edge and the Ramsey pulses at a fixed position relative to the readout pulse. 

The readout signal as a function of pulse duration is shown in Supplementary Figure~\ref{S_fig:FastFluxPulse}(b) for several values of pulse amplitudes. For short pulse duration, the pulse starts after the Ramsey sequence and does not change the readout signal. As the pulse duration increases, the leading edge of the pulse moves past the second and then the first $\pi/2$ pulse. The flux pulse experienced by the qubit in between the two $\pi/2$ pulses changes the phase of the Ramsey oscillation, causing a sharp change in the measured signal, with width corresponding to $\tau_\text{delay}$. The Ramsey signal shown in Supplementary Figure~\ref{S_fig:FastFluxPulse}(b) is flat up to flux pulse duration $\tau_\text{flux}$ as long as 30 $\mu s$, indicating that the pulse is negligibly distorted at these time scales. 
\begin{figure}
    \centering
    \includegraphics{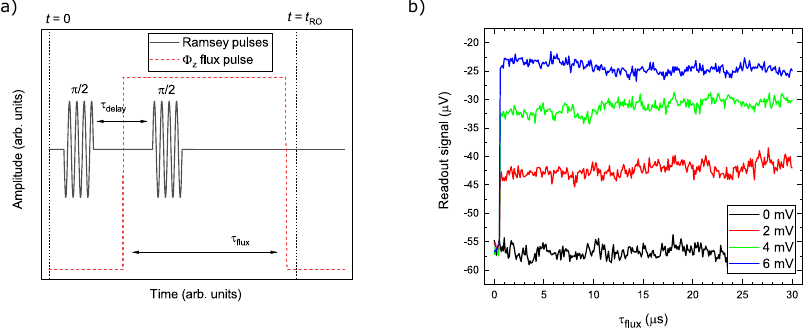}
    \caption{Characterization of pulse distortion from the bias tee. a) Schematic of Ramsey-based sequence used to quantify the pulse distortion. b) Readout voltage as a function of pulse duration $\tau_{\text{flux}}$. The flat profile of the readout voltages out to several tens of $\mu$s shows that the bias tee negligibly distorts the pulse shape. }
    \label{S_fig:FastFluxPulse}
\end{figure}

\section{State preparation and readout calibration}\label{S_sec:StatePrepReadout}
\begin{figure}
    \centering
    \includegraphics{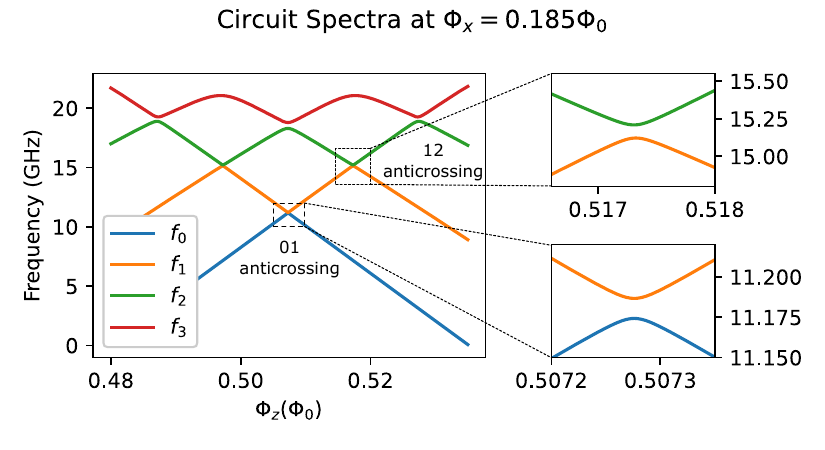}
    \caption{Energy Spectrum of the qubit circuit at $\Phi_x=0.185\Phi_0$, with close up of the 01 and 12 anticrossing}
    \label{S_fig:Energy}
\end{figure}
In this section, we describe the protocol for preparing the qubit in its ground state and the method for calibrating the readout voltage to obtain the state populations.

The qubit can in general be prepared in the ground state by waiting long enough. However, for some of the small gap $\Delta$ values used in the Landau-
Zener experiments, the transition rate between the ground states of the two wells is very slow. Therefore, to prepare the qubit in the ground state before the Landau-Zener measurements, we use a  sideband cooling method similar to that used in~Ref.~\cite{valenzuela_2006_microwaveinducedcoolingsuperconducting}
A plot of the qubit energy levels is shown in Supplementary Figure~\ref{S_fig:Energy}. The qubit is prepared in its ground state at a bias about $\Phi_z -\Phi_{z,\text{sym}}\approx 0.005$ away from the symmetry point, as follows. 
A sinusoidal pulse is applied which, on its positive side, sweeps the qubit further away from the symmetry point and past the anti-crossing between the first and second excited states. Prior to this sinusoidal pulse, the qubit is in a mixture of the ground and excited states. Due to the larger anti-crossing between the higher levels, the excited state can be adiabatically transferred into the same well as the ground state, which allows fast relaxation back to the ground state. By repeated sweeping across the anti-crossing between the first and second excited state, it is possible to prepare the qubit in the ground state. In our experiments, the qubit was cycled 5-10 times in order to prepare the ground state prior to the Landau-Zener sweep.
  
Next, we discuss the calibration of the readout signal. Given a fixed readout frequency, the transmission measured is given by a complex number $V_g (V_e)$ for the qubit at the ground (excited) state. For a qubit in the mixed state, the readout voltage is given by $V=V_g+P_e(V_e-V_g)$ where we assume populations beyond the qubit states are negligible. Therefore to obtain the qubit excited state population we need to obtain $V_g$ and $V_e$.

For the Landau-Zener measurements, $V_g$ and $V_e$ are calibrated at each $\Phi_x$. To measure $V_g$ we prepare the qubit in the ground state at the readout point (at the end of the Landau-Zener sweep) using the cooling procedure discussed above. To measure $V_e$, we prepare the qubit in the excited state by preparation in the ground state at the opposite side of the symmetry point, followed by a fast, 1~ns long, Landau-Zener sweep through the minimum gap. Taking the qubit model parameters and the coherent Landau-Zener formula, a 1~ns Landau-Zener is expected to lead to a final excited state probability larger than 99\% for even the largest $\Delta$ measured in our experiments.  

\section{Noise parameters and master equation simulation}\label{S_sec:NoiseParameters}
In this section, we discuss the noise model used in the master equation simulation. We first introduce the general form of noise and then discuss some specificities regarding including them into the adiabatic master equation (AME) and polaron-transformed Redfield equation (PTRE). For reference, we also provide a comparison of the noise parameters used in this work with noises measured in three other flux-qubit based quantum annealing devices, as summarized in Supplementary Table~\ref{S_tab:NoiseParameters}.

As the qubit circuit has relatively large flux loops, we assume the flux noise in the qubit $x$ and $z$ loops are the dominant sources of noise. The noises lead to fluctuation in the circuit Hamiltonian via
\begin{align}
    \delta H_{\t{c}}(\Phi_z, \Phi_x) = \sum_{\lambda\in\{\Phi_z,\Phi_x\}}
    \frac{\partial H_{\t{c}}}{\partial \lambda}\delta \lambda.
\end{align}
In particular for $\lambda=\Phi_z$, in the two level approximation $\frac{\partial H_{\t{c}}}{\partial \Phi_z}=-I_p\sigma_z$. Note here when the noise source is quantum, $\delta \lambda$ is a quantum operator of the environment.

The noise power spectral density (PSD) due to $\delta \lambda$ is given by the Fourier transform of its auto-correlation function, 
\begin{align}
    S_{\lambda}(\omega) = \int^\infty_{-\infty}\text{d}\tau e^{i\omega \tau}\langle \delta\lambda(\tau) \delta\lambda(0)\rangle.
\end{align}
Various previous measurements have shown that flux noise has roughly $1/f$ dependence, with $f$ being frequency, up to around 1GHz, and then a quasi-ohmic spectrum at higher frequencies~\cite{yan_2016_fluxqubitrevisited,quintana_2017_observationclassicalquantumcrossover,lanting_2011_probinghighfrequencynoise}. Furthermore, we follow Ref.~\cite{quintana_2017_observationclassicalquantumcrossover} to consider a quantum noise PSD with the positive and negative frequency components related by a phenomenological thermodynamic model. These considerations lead to the noise PSD given by
\begin{align}
    S_{\lambda} &= S_{\lambda,\text{1/f}} + S_{\lambda,\text{ohmic}}\label{S_eq:NoisePSD}\\
    S_{\lambda,\text{1/f}}&= \frac{A_{\lambda}\omega}{|\omega|^\alpha}\l[1+\coth\l(\frac{\beta\hbar\omega}{2}\r)\r],\\
    S_{\lambda,\text{ohmic}}&=B_{\lambda}\omega|\omega|^{\gamma-1}\l[1+\coth\l(\frac{\beta\hbar\omega}{2}\r)\r],
\end{align}
where $\beta=1/k_B T$ is the inverse temperature, $A_\lambda, B_\lambda$ determines the noise strength and $\alpha, \gamma$ determine the frequency dependence. For $1/f$ noise, $\alpha= 1$ and for ohmic noise $\gamma=1$. The temperature is assumed to be close to the base temperature of the fridge, $T=20\,\t{mK}$.

The parameters of $1/f$ noise are obtained via measuring the flux-bias dependent Ramsey dephasing times (see Supplementary Figure~\ref{fig:T1Tphi}). The Ramsey dephasing time probes symmetrized $1/f$ noise in the low-frequency limit, satisfying $\hbar \omega\ll k_B T$. In this limit, we have

\begin{align}
    S^+_{\lambda,1/f}(\omega)&=\frac{1}{2}\l(S_{\lambda,1/f}(\omega)+S_{\lambda,1/f}(-\omega)\r)\\
    &\approx \frac{A_{\lambda}\omega}{|\omega|^\alpha}\frac{2}{\hbar\omega\beta}\\
    &=A^*_{\lambda}\left(\frac{2\pi}{|\omega|}\right)^\alpha,
\end{align}
where we defined $A_\lambda^*=2A_\lambda/[\hbar\beta(2\pi)^\alpha]$ to relate to the more commonly used expression for $1/f$ flux noise, used in for example Ref.~\cite{weber_2017_coherentcoupledqubits}. Given the similarity between our device and the device used in  Ref.~\cite{weber_2017_coherentcoupledqubits}, we assume $\alpha=0.91$ and found $A_{\Phi_z}^*=\l(8.7\times 10^{-6}\r)^2\Phi_0^2/\t{Hz}$, $A_{\Phi_x}^*=\l(5.\times 10^{-6}\r)^2\Phi_0^2/\t{Hz}$ fits the measured Ramsey dephasing time best. More details on the coherence characterization of this qubit are discussed in a separate publication~\cite{trappen_2023_decoherencetunablecapacitively}. 
\begin{figure}
    \centering
    \includegraphics{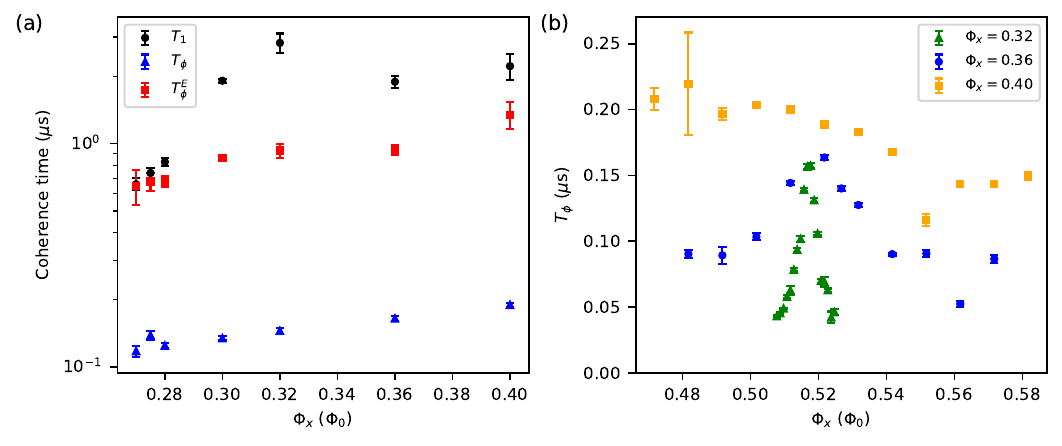}
    \caption{Measured coherence times of the device. a) The measured relaxation time~($T_1$), Ramsey and spin-echo pure dephasing time~($T_\phi, T_\phi^E)$ at the symmetry points of $\Phi_z$ for different $\Phi_x$. b) The measured Ramsey pure dephasing time $T_\phi$ at different $\Phi_z$ for three different $\Phi_x$ values. The measured coherence times are used to find the noise parameters used in the master equation simulations.}
    \label{fig:T1Tphi}
\end{figure}
The quasi-ohmic component of the flux noise mainly contributes to qubit relaxation. It is more difficult to give a quantitative estimate of the quasi-ohmic noise power as it leads to similar flux bias dependence of relaxation rates as other noise sources, such as ohmic charge noise~\cite{yan_2016_fluxqubitrevisited,quintana_2017_observationclassicalquantumcrossover}. For this reason, we use the reported ohmic noise strength measured in Ref.~\cite{quintana_2017_observationclassicalquantumcrossover} and scale it according to the ratio of the $1/f$ noise strength at 1Hz between the two devices. This gives $\gamma=1$ and $B_{\Phi_z}\approx2.7\times 10^{-30}\Phi_0^2/\t{Hz}^2, B_{\Phi_x}\approx 9.2\times 10^{-31}\Phi_0^2/\t{Hz}^2$. These values give reasonable agreement with the qubit $T_1$ relaxation times we have measured. In Supplementary Table~\ref{S_tab:NoiseParameters}, we also provide these numbers in terms of the dimensionless coupling constant $\eta$ that is often discussed in spin-boson literature. We also want to note that the simulation result is largely unchanged if the ohmic component of the noise spectrum is not included. 

\begin{table}
\begin{ruledtabular}
\begin{tabular}{c|c|c|c|c}
     & This work & Quintana et. al~\footnote{The numbers are based on Ref.~\cite{quintana_2017_observationclassicalquantumcrossover}}& DWave CJJ qubit~\footnote{The numbers are based on Ref.~\cite{johansson_2009_landauzenertransitionssuperconducting} where the Landau-Zener experiment is performed.} &DWave CCJJ qubit~\footnote{The numbers are based on Ref.~\cite{harris_2010_experimentaldemonstrationrobust,lanting_2011_probinghighfrequencynoise}}\\
     
     $A_{\Phi_z}^*$ ($\Phi_0^2/\text{Hz})$& $(8.7\times10^{-6})^2$ & $\sim (5\times10^{-6})^2/2$ &&$(1.3\times10^{-6})^2/2$\\
     
     $\alpha$ & $0.91$&$0.96 - 1.05$&&$0.95$\\
     
     $I_p$($\mu$A)&$0.104-0.129$ &$\sim0.5$&&$\sim 1.0$ \\
     
     $W/h$(predicted)(GHz) & $0.048-0.059$& $\sim 0.16$&&$0.05$\\
     
     $W/h$(measured)(GHz) & &$\sim 0.25$&$2.6$&$1.4$\\
     
     $T_{\t{eff}}(\t{mK})$\footnote{This is the effective temperature that describes the MRT data assuming low-frequency noise is at thermal equilibrium.}&$20$&$20$&$21$&$53$\\
     
     $\eta g^2$\footnote{$\eta g^2$ is defined such that, the system bath coupling is given as $g A\otimes B$, where $A, B$ are norm-1 system and bath operators respectively and the noise PSD of $B$ is given as $S(\omega)=2\pi\eta \hbar^2\omega \frac{\exp(-|\omega|/\omega_c)}{1-\exp(-\beta\hbar\omega)}$. See for example Ref.~\cite{albash_2012_quantumadiabaticmarkovian} for more details.}&$1.4\times 10^{-5}$&$\sim 5\times10^{-5}$&&$\sim0.065$\cite{lanting_2011_probinghighfrequencynoise}
\end{tabular}
\end{ruledtabular}
\caption{Comparison of noise parameters used for simulation in this work and other work using flux qubits for quantum annealing.}
\label{S_tab:NoiseParameters}
\end{table}

\subsection{Effect of X-noise coupling}\label{S_sec:XNoise}
In the master equation simulations, noise from $\Phi_x$ is not included. This is justified based on three considerations. Firstly, the noise power of $\Phi_x$ is less than half the noise power in $\Phi_z$. Secondly, as plotted in Supplementary Figure~\ref{S_fig:FluxMatrixElements}, the matrix elements of the flux operators between the circuit energy eigenstates $\langle \alpha|\partial H/\partial \Phi_x|\beta\rangle$ is smaller than $\langle \alpha|\partial H/\partial \Phi_z|\beta\rangle$ by a factor of 10. Finally, $\Phi_x$ noise only primarily leads to transverse noise assuming small $\Phi_x$ and small $x$ loop junction asymmetry. Previous studies suggested that for transverse noise to have similar dissipative effects, its coupling strength needs to be at least about 1/10 of the longitudinal noise~\cite{javanbakht_2015_dissipativelandauzenerquantum}; this condition is far from being satisfied in our case. 

\begin{figure}
    \centering
    \includegraphics[]{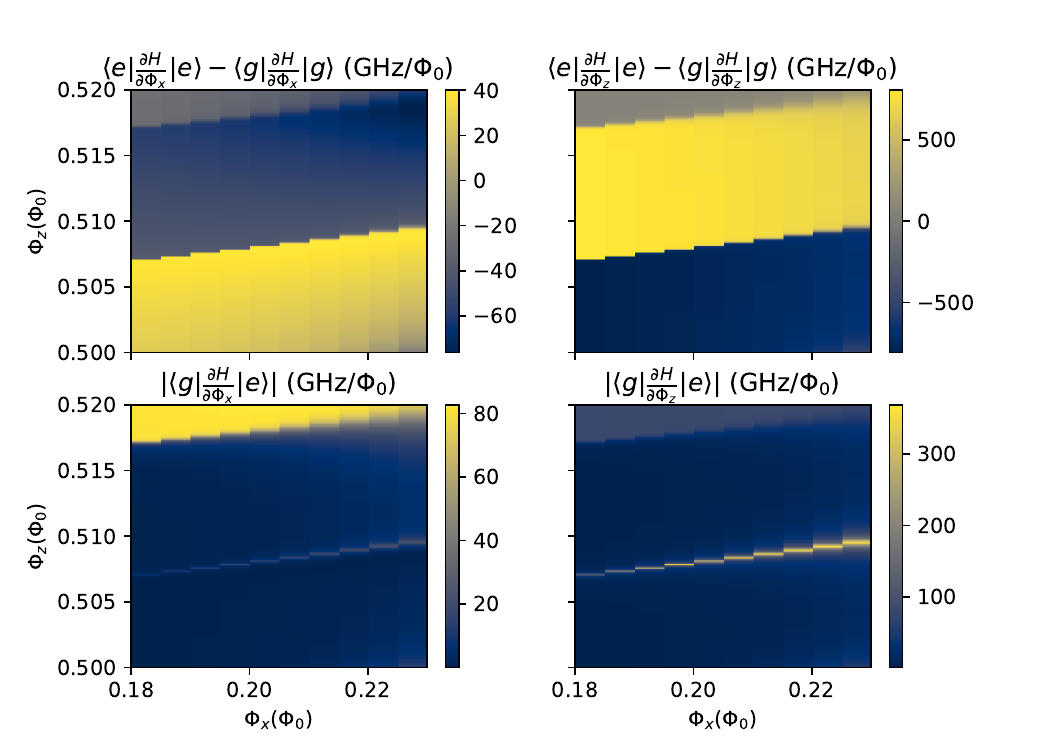}
    \caption{Matrix elements for the $x$ and $z$ flux bias. }
    \label{S_fig:FluxMatrixElements}
\end{figure}

\subsection{Adiabatic master equation (AME)}\label{S_sec:AME}
In order to implement in AME the noise PSD in Supplementary Eq.~(\ref{S_eq:NoisePSD}), additional low- and high-frequency cutoffs $\omega_l$ and $\omega_h$ are added to avoid divergence. This introduces $S_{\Phi_z}^{\t{AME}}$, the PSD used in AME simulations, defined as 

\begin{align}
    S_{\Phi_z}^{\t{AME}}&=\begin{cases}
    S_{\Phi_z,\t{1/f}}(\omega)\exp\l(\frac{-|\omega|}{\omega_h}\r)+
    S_{\Phi_z,\t{ohmic}}(\omega)\exp\l(\frac{-|\omega|}{\omega_h}\r)
    &|\omega|>\omega_l\\
    S_{\Phi_z,\t{1/f}}(\omega_l)\exp\l(\frac{-|\omega_l|}{\omega_h}\r)+
    S_{\Phi_z,\t{ohmic}}(\omega)\exp\l(\frac{-|\omega|}{\omega_h}\r) &|\omega|\leq\omega_l.
    %   \frac{2A_\t{\Phi_z}^2}{(|\omega|/2\pi)^\alpha}\frac{1}{1 + \exp(-\hbar\omega/k_B T)},& \omega_l <|\omega| \leq \omega_h \\
    %   \frac{2A^2}{(\omega_l/2\pi)^\alpha}\frac{1}{1 + \exp(-\hbar\omega/k_B T)}, & 0 <\omega \leq \omega_l \\
    %   \frac{2A^2}{(\omega_l/2\pi)^\alpha}\frac{1}{1 + \exp(-\hbar\omega/k_B T)}, & -\omega_l <\omega \leq 0 \\
    %   0, & \text{otherwise},
    \end{cases}
\end{align}
The high-frequency cutoff is chosen to be $\omega_h/2\pi=10\t{GHz}$, which is roughly the characteristic oscillation frequency in either of the qubit potential wells. For the low-frequency cutoff, given that we are primarily concerned with thermalization effects, we choose $\omega_l/2\pi$= $10\t{MHz}$, which corresponds to the minimum qubit frequency for the Landau-Zener measurement presented in this work.

\subsection{Polaron-transformed master equation (PTRE)}\label{S_sec:PTRE}
In the PTRE simulation, the noise model consists of an ohmic noise just as in AME, and the $1/f$ noise is represented by the MRT parameters $W$ and $\epsilon_p$~\cite{amin_2008_macroscopicresonanttunneling}. The MRT width $W$ characterizes the integrated effect of the symmetrized low-frequency noise,

\begin{align}
    W^2 = 2I_p^2\int_{\omega_{\t{low}}}^{\omega_{\t{high}}}\frac{d\omega}{2\pi} S_{\Phi_z,\t{1/f}}^+(\omega).\label{eq:MRTWidth}
\end{align}

The anti-symmetrized low frequency noise $S_{\Phi_z,\t{1/f}}^-(\omega)=1/2(S_{\Phi_z,\t{1/f}}(\omega)-S_{\Phi_z,\t{1/f}}(-\omega))$ gives the reorganization energy $\epsilon_p$
\begin{align}
    \epsilon_p=2I_p^2\int_{\omega_{\t{low}}}^{\omega_{\t{high}}}\frac{d\omega}{2\pi}\frac{S_{\Phi_z,\t{1/f}}^-(\omega)}{\hbar\omega}.
\end{align}
For the integration limit, we choose $\omega_{\t{low}}/2\pi=4\,\t{Hz}$ based on the experiment time taken for all the repetitions at each $\Phi_x$ and $T_{\text{LZ}}$, and $\omega_{\t{high}}/2\pi=10\,\t{Ghz}$ based on the characteristic oscillation frequency in the qubit potential wells. We assume that the low-frequency noise is in thermal equilibrium, which relates $W$ and $\epsilon_p$ via the fluctuation-dissipation theorem, $W^2=2k_BT\epsilon_p$. We also note that $1/f$ noise has a significant contribution to the noise power at high frequency, up to around $1\,\text{GHz}$. This contribution breaks the normalization condition for the high-frequency noise in the current numerical implementation of PTRE (see discussion around Eq.~[16] of Ref.~\cite{smirnov_2018_theoryopenquantuma}). The effect of the high-frequency component of $1/f$ noise in the strong coupling limit is to be explored in future work. 

As discussed in the main text, the values of $W$ or $T$ need to be adjusted for better agreement between PTRE simulation and the experiment data at low $\Phi_x$. The results are shown in Supplementary Figure~\ref{S_fig:PTRE}. Interestingly, it is found that increasing $W$ by 4(8) times is equivalent to decreasing $T$ by 4(8) times. This indicates that the ratio $\epsilon/W$ is most critical to the result, with increasing $\epsilon_p/W$ leading to closer to the coherent limit of ground state population. 

\begin{figure}
    \centering
    \includegraphics[scale=0.5]{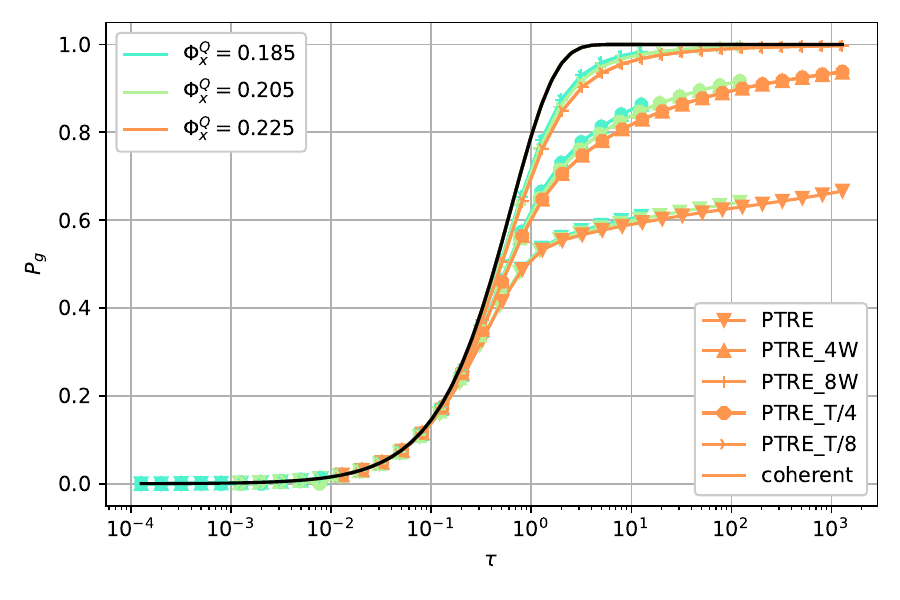}
    \caption{Final ground state probabilities versus the dimensionless sweep time $\tau=\Delta^2/\hbar v$ for different PTRE parameters.}
    \label{S_fig:PTRE}
\end{figure}
\subsection{Symmetric versus Asymmetric Landau-Zener Sweep}\label{S_sec:AsymmetricSweep}
As discussed in \ref{S_sec:CircuitModel}, the screening current in the rf-SQUID leads to an effective bias to the qubit $z$ loop. Due to an initial inaccurate estimation of this effect, the Landau-Zener data presented in the main text has an asymmetric scan range, with initial and final $z$ loop bias being $\Phi_{z,\t{init}}=-3.1\times10^{-3}\Phi_0$ and $\Phi_{z,\t{final}}=6.9\times10^{-3}\Phi_0$. This was later identified via the spectroscopy method mentioned in \ref{S_sec:CircuitModel}, but leaving insufficient time to repeat the full range of Landau-Zener experiments. However, with this asymmetric scan range, the validity of the Landau-Zener model is not affected, since the initial and final longitudinal field are still much larger than the tunneling amplitude. In Supplementary Figure~\ref{S_fig:AsymLZ}, we compare the measured and simulated results for symmetric and asymmetric $\Phi_z$ sweep range. The experiment data show some differences but the qualitative features discussed in the main text remain the same. In particular the symmetric data also show that as $\Phi_x$ decreases, $P_g$ becomes closer to the coherent limit behaviour. The simulated results using either AME or PTRE for the symmetric and asymmetric sweep range do not differ significantly. 

\begin{figure}
    \centering
    \includegraphics[scale=0.6]{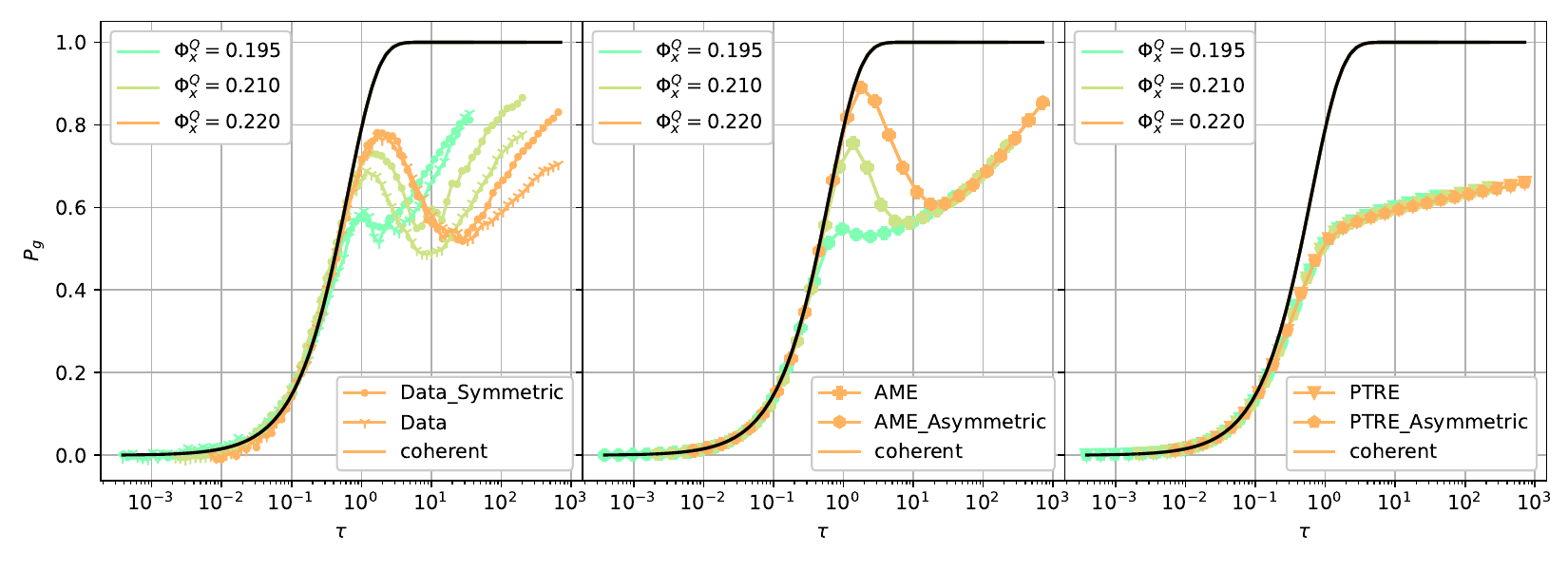}
    \caption{Final ground state probabilities comparing the symmetric and asymmetric sweep around $\Phi_z$ symmetry point, for experiment data on the left panel, AME simulation on the center panel and PTRE simulation on the right panel. The simulated results uses the nominal noise parameters discussed in \ref{S_sec:NoiseParameters} and the symmetric and asymmetric sweep overlap. The experiment data presented in the main text is asymmetric in range and the simulated data in the main text is symmetric in range.}
    \label{S_fig:AsymLZ}
\end{figure}

\section{Anomalous Population inversion at $\Phi_x=0.2$}
As noted in the main text, in the Landau-Zener measurement, the final ground state probability, $P_g$ for $\Phi_x=0.2$, drops below 0.5 for intermediate sweep time. This is inconsistent with the interpretation that the dominant mechanism determining $P_g$ at intermediate sweep time is thermalization that occurs near the minimum gap. A possible explanation is imperfect ground state preparation just before the Landau-Zener sweep, which affects both the Landau-Zener measurement itself and the calibrated $V_e$. Assuming the ground state initialization procedure leaves a finite population in the initial excited state, the result of the Landau-Zener measurement is shown in Supplementary Figure~\ref{S_fig:ReadoutError}(a), where we have denoted the actual~(miscalibrated) excited state voltage as $V_e~(V_e^\prime)$, and the corresponding inferred final ground state probability as $P_e~(P_e^\prime)$. We found that an initial excited state population $\sim 0.1$ is needed to eliminate the anomalous population inversion at intermediate sweep time. 
\begin{figure}
    \centering
    \includegraphics[width=7in]{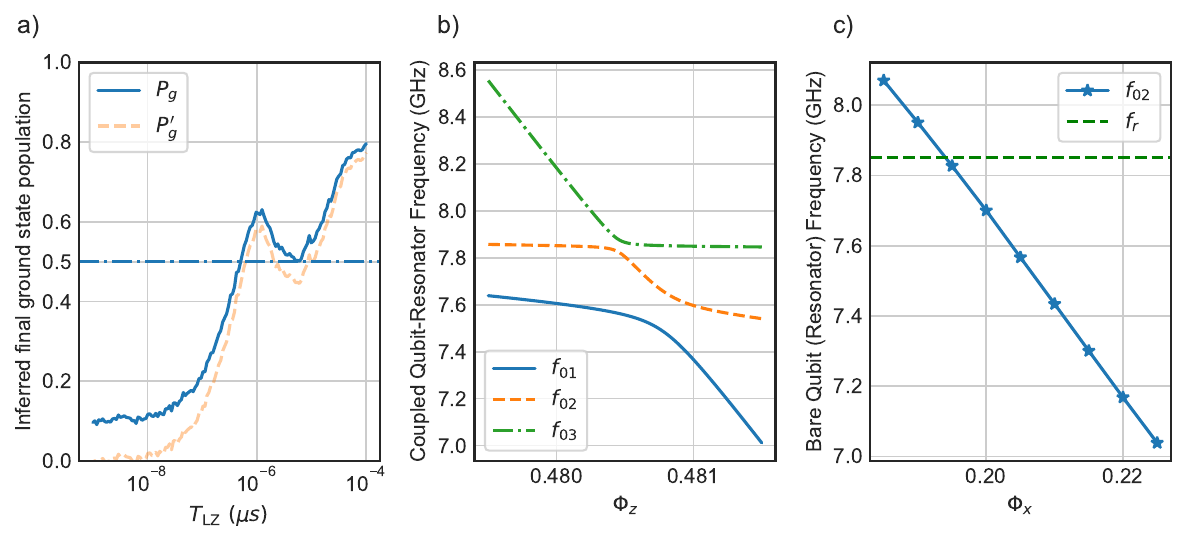}
    \caption{a) Inferred final ground state probabilities of the Landau-Zener measurement versus the Landau-Zener sweep time for $\Phi_x=0.2$, $P_g$ (blue solid line) assuming 0.1 error in ground state initialization, and $P_g^\prime$ (faint orange dashed line) assuming perfect ground state initialization. The data shown in the main text is $P_g^\prime$. b) Numerically simulated transition frequencies of the coupled qubit-resonator system for $\Phi_x=0.2$, near the 1-2 anti-crossing (between blue solid line and orange dashed lines) used to initialize the qubit. The additional anti-crossing between the orange dashed line and green dot-dashed lines indicates anti-crossing between the resonator and the qubit's second excited state. c) The qubit's 0-2 transition frequency versus $\Phi_x$, with $\Phi_z$ chosen to be right at the 1-2 anti-crossing. The frequency collision between the qubit's 0-2 transition frequency and the resonator does not occur exactly around 0.2, which could be due to imperfections in the circuit modeling of the multi-level system.}
    \label{S_fig:ReadoutError}
\end{figure}

The source of imperfect ground state preparation at this particular $\Phi_x$ value is likely due to frequency collision in the coupled qubit-resonator system. In Supplementary Figure~\ref{S_fig:ReadoutError}~(b, c), we show the bare resonator frequency $f_r$ and the bare qubit ground to second excited state transition frequency $f_{02}$ versus the qubit $x$ flux bias $\Phi_x$. It can be seen that at a certain $\Phi_x$ value close to $0.2$, $f_{02}$ crosses the resonator frequency $f_r$. Due to finite interaction strength between the qubit and the resonator (estimated to be about 30~$\mathrm{MHz}$), the frequency collision opens another anti-crossing, which is close to the anti-crossing between the first and the second excited state of the qubit circuit. When the two anti-crossings are close, the system dynamics going through them are considerably more complex than the two-level picture, which could introduce errors in the ground state preparation. However, at other $\Phi_x$ values, when the two anti-crossings can be treated separately when they are far apart (when $f_{02}$ is far from $f_r$). In this case, any population transfer to the resonator would also quickly decay to the collective ground state of the qubit-resonator system, aiding the cooling procedure.

\section{Spin bath}\label{S_sec:SpinBath}
Given that the Markovian master equations, AME and PTRE, failed to capture the crossover from the weak to strong coupling limit in the experiment data, it is natural to ask whether this crossover can be captured by simulating the experiment incorporating an explicit quantum environment. The spin bath is a natural choice for this quantum environment. First, much theoretical and experimental evidence points to spin impurities being the source of $1/f$ flux noise. Second, spins ferromagnetically coupled to the qubit offer an intuitive picture of the MRT phenomenon, a canonical example of the strong coupling limit. The non-zero expectation of the spins' polarization acts as an environmental bias to the qubit, resembling the reorganization energy in MRT. Fluctuations of this polarization due to the internal dynamics of the spins effectively dephase and broaden the longitudinal bias seen by the qubit, analogous to the MRT tunneling width. 
\subsection{Single spin bath}
The simplest toy model for a quantum environment coupled to the system is a single spin ferromagnetically coupled to the system qubit, with an additional bath that causes thermalization of the spin. This system can be described by the Hamiltonian
\begin{align}
    H &= H_q + H_{qb} + H_{qS}+ H_{SB} + H_S  + H_b + H_{B}\\
    H_{qb} &= I_p\sigma_zQ_{\Phi_z}\\
    H_{qS} &= J\sigma_z\tau_{z}\\
    H_S &= 0\\
    H_{SB} &= \tau_{x} Q^\prime\\
    S_{Q_{\Phi_z}}(\omega)&=S^\text{AME}_{\Phi_z}(\omega)\\    
    S_{Q^\prime}(\omega)&=\hbar^2\lambda\frac{1}{1+\exp{(-\beta\hbar\omega)}}\exp{\left(-\frac{\omega}{\omega_c}\right)},
\end{align}
The model Hamiltonian is understood as follows. First, the high-frequency, Markovian part of the noise is captured by the noise PSD used in the previous AME simulation. On top of this, a single spin is added to capture the effect of strong low-frequency noise. The qubit is ferromagnetically coupled to the spin with coupling strength $J$. The environmental spin does not have internal dynamics, but it is transversely coupled to its own environment. This environment is nearly white noise, but with the induced relaxation and excitation rate of the spin satisfying detailed balance. The strength of this noise $\lambda$ is essentially the thermalization rate of the spin, in the limit where the qubit-spin coupling approaches zero. Similar to the noise PSD describing the environment of the qubit, the noise PSD for the environment of the spin also has an exponential cutoff, the role of which is essentially to ensure numerical convergence.  

Next, assuming the temperature and the cutoff frequency of the spin's environment are the same as the qubit's environment, there are only two parameters to the model, the coupling strength $J$ and the free spin thermalization rate $\lambda$. We numerically simulate the Landau-Zener experiment with the above model, initializing the system in the ground state (initialization in the thermal state does not significantly change the result). The result is shown in Supplementary Figure~\ref{fig:SingleSpinBath}, with a range of $J$ and $\lambda$ taken in geometric steps.
\begin{figure}
    \centering
    \includegraphics{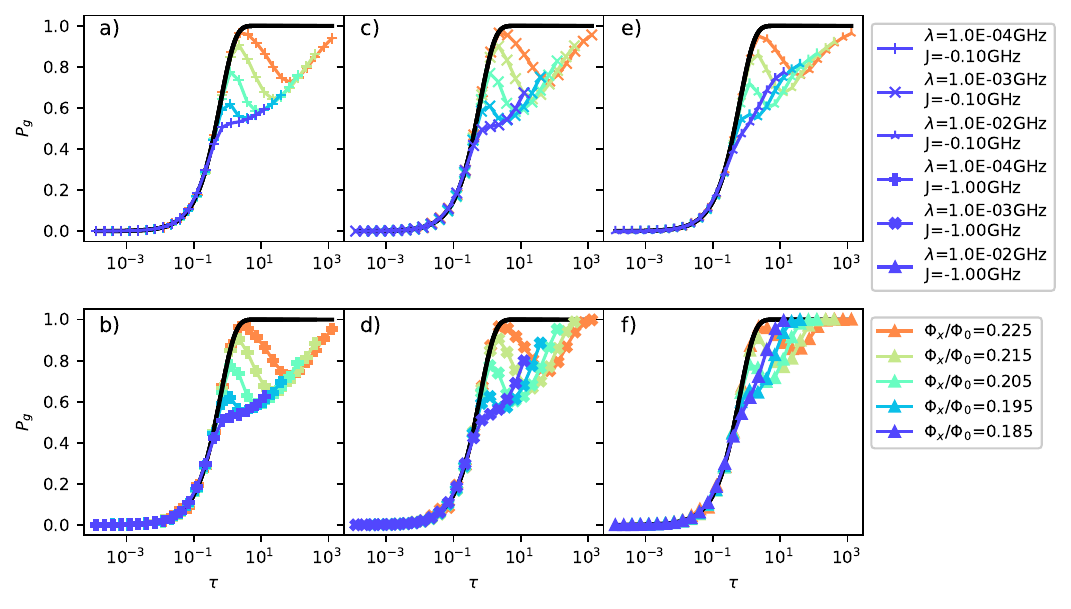}
    \caption{Final ground state probabilities versus the dimensionless sweep time $\tau$, using the single spin bath model (see text). Parameters for the single spin are $\lambda=0.0001, 0.001, 0.01$ GHz from left to right and $J/h=-0.1 (1.0)$ GHz on the top(bottom).}    \label{fig:SingleSpinBath}
\end{figure}
It can be seen that for small $J$ and $\lambda$, the result of the single spin bath model is almost the same as the qubit AME result. This can be understood from two perspectives. First, small $\lambda$ corresponds to the spin being nearly coherent and having no dynamics. Therefore its effect is to merely shift the location of the anticrossing to $\epsilon+2J=0$. Second, for small $J$, the thermal equilibrium state of the spin always has its polarization $\langle\tau_z\rangle\approx 0$, irrespective of the qubit state. Therefore it has a negligible impact on the qubit. As $\lambda$ and $J$ increase, the spin bath model predicts a behavior that closely resembles the experimental data. For large $\Phi_x$, the single spin bath model simulated ground state probably has a non-monotonic dependence versus Landau-Zener sweep time, but as $\Phi_x$ decreases, the dependence becomes monotonic and approaches the coherent limit. Among the parameters simulated, there is a good qualitative agreement between the simulated and experiment data for $\lambda=0.001$ GHz and $J/h=1.0$ GHz. 

Further insight into the model can be obtained by looking at the instantaneous state probabilities, which are shown in Supplementary Figure~\ref{S_fig:SingleSpinEvolve}. First, when $\lambda$ is small (Supplementary Figure~\ref{S_fig:SingleSpinEvolve}(a)), the spin is nearly coherent and simply adds an additional bias $J$ to the qubit. This shifts the position of the anti-crossing, but does not change the final ground state probabilities, as long as $J$ is well within the initial and final $Z$ bias of the qubit. For small $J$ such that $\beta |J|\ll 1$ (Supplementary Figure~\ref{S_fig:SingleSpinEvolve}(b)), the thermal average of the spin's polarization is almost always zero, regardless of the state of the qubit, or the thermalization rate $\lambda$, Therefore the spin has a negligible effect on the qubit.
\begin{figure}
    \centering
    \includegraphics{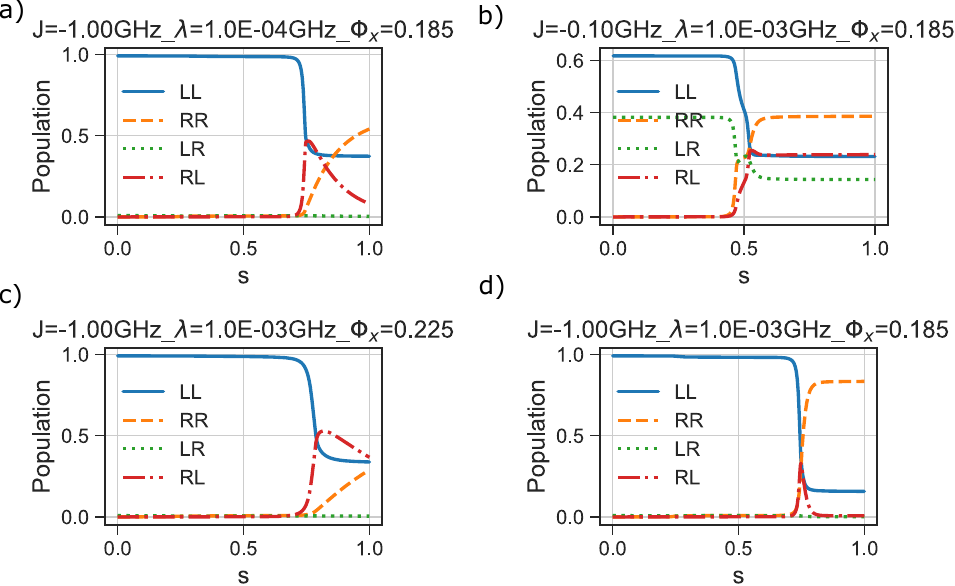}
    \caption{Evolution of the instantaneous qubit and spin states as a function of the normalized time $s\in[0, 1]$, for different parameters of the qubit and the spin.}
    \label{S_fig:SingleSpinEvolve}
\end{figure}

Finally, for $\beta J\gtrsim 1$, qualitative differences arise for large and small $\Delta$ (Supplementary Figure~\ref{S_fig:SingleSpinEvolve} (c) and (d) respectively). When $\Delta$ is large, the qubit completes the tunneling before the spin has time to re-align with the qubit. This allows thermalization to happen across the symmetrized and anti-symmetrized states of the qubit, which is why the ground state probability has a local minimum of $P_g\approx0.5$. However, when $\Delta$ is small, the spin quickly relaxes to the opposite state as well, following the qubit. After the spin has relaxed, the qubit effectively sees a $Z$ bias of $-2J$, which suppresses thermalization induced by the qubit's own environment. In other words, the relatively fast relaxation of the spin makes qubit tunneling irreversible. Therefore for small $\Delta$, as the sweep time increases, the ground state probability no longer has a local minimum around $0.5$.

\subsection{Multiple spins}
A general spin bath has a large parameter space, such as the distribution of their longitudinal and transverse fields, the distribution of couplings among them, and the distribution of couplings to their respective environments. While some of these parameters can be motivated based on plausible models of flux noise, these models often assume a macroscopic number of spins, and it is not clear whether they have an efficient numerical representation. Given this complexity, we leave it as future work to systematically explore the parameters of the spin bath and its relation to the physical $1/f$ flux noise. In this work, we adopt a simplified approach which can be seen as a quantum extension of simulating classical $1/f$ noise with two-level fluctuators~\cite{dutta_1981_lowfrequencyfluctuationssolids}. 

In the multi-spin model, the Hamiltonian changes from the single-spin model by making the changes given by
\begin{align}
    H_{qS} &= \sum_i^{N_s}J_i\sigma_z\tau_{z,i},\\
    H_{SB} &= \sum_i^{N_s}\tau_{x,i} Q_i^\prime,~\text{and}\\
    S_{Q_i^\prime}(\omega)&=\hbar^2\lambda_i\frac{1}{1+\exp{(-\beta\hbar\omega)}}\exp{\left(-\frac{\omega}{\omega_c}\right)}.
\end{align}
The spin bath parameters $J_i$ and $\lambda_i$ can be chosen based on the measured $1/f$ flux noise strength. To see this, we first consider the large $N_s$, weak coupling limit, where the effect of the spin bath can be well captured by the noise PSD. Following Ref.~\cite{shnirman_2005_lowhighfrequencynoise}, the symmetrized noise PSD of the $i$'th spin is
\begin{align}
    S_i(\omega)&=(1 -\langle\tau_i\rangle)\frac{2\gamma_i J_i^2}{\omega^2+\gamma_i^2},\label{eq:SpinPSD}
\end{align}
where $\langle\tau_i\rangle$ is the expectation value of spin $i$'s longitudinal polarization, and $\gamma_i$ is its thermalization rate. In general, the expectation value and the thermalization rate depend on the instantaneous qubit Hamiltonian. However, for a small enough coupling between the qubit and the spin, the effect of the qubit on the spin is an effective longitudinal bias with strength, $-J_i\langle \sigma_z\rangle/2$. In this case, the expectation value and the thermalization rate are
\begin{align}
    \langle \tau_i\rangle &=\tanh{\left(\beta\langle\sigma_z\rangle J_i\right)},\\
    \gamma_i&=\lambda_i\exp\left(-\beta|J_i|\langle\sigma_z\rangle\right).
\end{align}
Then to obtain $1/f$ like noise with exponent $\alpha$, we assume that the distribution of $\lambda$ for all the spins is given by
\begin{align}
    P_\lambda(\lambda)&=\frac{1}{N_\lambda}\frac{1}{\lambda^\alpha},\\
    N_\lambda&=\left(\frac{1}{-\alpha+1}\right)\left(\lambda_\text{max}^{-\alpha+1}-\lambda_\text{min}^{-\alpha+1}\right),
\end{align}
where $\lambda_\text{min}, \lambda_\text{max}$ can be chosen based on the frequency range of the noise PSD that we are interested in simulating. For simplicity, we also assume that the ferromagnetic coupling is constant for all spins, $J_i=J$. Then the collective noise PSD of the spin bath is 
\begin{align}
    S_S(\omega)&=N_s(1-\langle\tau\rangle)\int_{\lambda_\text{min}}^{\lambda_\text{max}}P_\lambda(\lambda)\frac{2\gamma_i(\lambda)J^2}{\omega^2+\gamma^2(\lambda)}\\
    &=(1-\langle\tau\rangle)\frac{N_sJ^2}{N_\lambda}c^{\alpha-1}\frac{1}{\omega^\alpha}\mathcal{I}
\end{align}
where we have introduced
\begin{align}
    \langle\tau\rangle&=\tanh{\left(\beta\langle\sigma_z\rangle J_i\right)},\\
    c&=\exp(-\beta|J_i|)~\text{and}\\
    \mathcal{I}&=\int_{c\lambda_\text{min}/\omega}^{c\lambda_\text{max}/\omega}\frac{x^{1-\alpha}}{1+x^2}dx.
\end{align} 
We can notice that for $\beta J\xrightarrow[]{}0$, the $J$ dependence of the noise PSD primarily comes from the $J^2$ term. This allows us to set $J$ based on the measured flux noise power. Comparing $S_S(\omega)$ with the symmetrized flux noise power we have
\begin{align}
    J=\sqrt{\frac{A_{\Phi_z}^*N_\lambda}{\mathcal{I}N_s}}I_p.
\end{align}
In the above expression, the integral $\mathcal{I}$ can be evaluated at a typical $\omega$ in between $\lambda_\text{max}$ and $\lambda_\text{min}$, and it is always close to $1$.

To confirm the intuition that the spin bath captures the MRT parameters, we can compare the following two expressions. First, from the fluctuation-dissipation theorem and the definition of MRT width using symmetrized noise PSD in Supplementary Eq.~\ref{eq:MRTWidth}, we have
\begin{align}
    \epsilon_p &= \frac{\beta}{2}W^2\\
    &=\beta I_p^2A_{\Phi_z}^*\frac{1}{-\alpha+1}\left[\left(\frac{\omega_\text{high}}{2\pi}\right)^{-\alpha+1}-\left(\frac{\omega_\text{low}}{2\pi}\right)^{-\alpha+1}\right].\label{eq:EpEstimate}
\end{align}
On the other hand, the effective bias applied to the qubit by the spin bath is
\begin{align}
    \epsilon_{SB}&=N_sJ\tanh{(\beta J)}\\
    &\approx N_s\beta J^2\\
    &=\beta A_{\Phi_z}^8I_p^2\frac{1}{\mathcal{I}}\frac{1}{-\alpha+1}\left[\left(\frac{\lambda_\text{high}}{2\pi}\right)^{-\alpha+1}-\left(\frac{\lambda_\text{low}}{2\pi}\right)^{-\alpha+1}\right].\label{eq:EffectiveBias}
\end{align}
Therefore, if we choose $\lambda_\text{max}\gtrsim \omega_\text{max}$ and $\lambda_\text{min}\lesssim \omega_\text{min}$, these two expressions, Supplementary Eq.~\ref{eq:EpEstimate} and \ref{eq:EffectiveBias} indeed match each other, up to a constant $\mathcal{I}$ that is close to unity.

As mentioned in the main text, when the low-frequency noise is not large enough, the spin bath simulation results closely resemble that of the single-qubit AME. This is exemplified in Supplementary Figure~\ref{fig:3SpinNominalAfz}, where the spin bath parameters are chosen to target an $1/f^\alpha$ noise spectrum with the same amplitude as deduced from the decoherence measurements, therefore being 8 times smaller than the simulation presented in the main text. 
\begin{figure}
    \centering
    \includegraphics[width=4in]{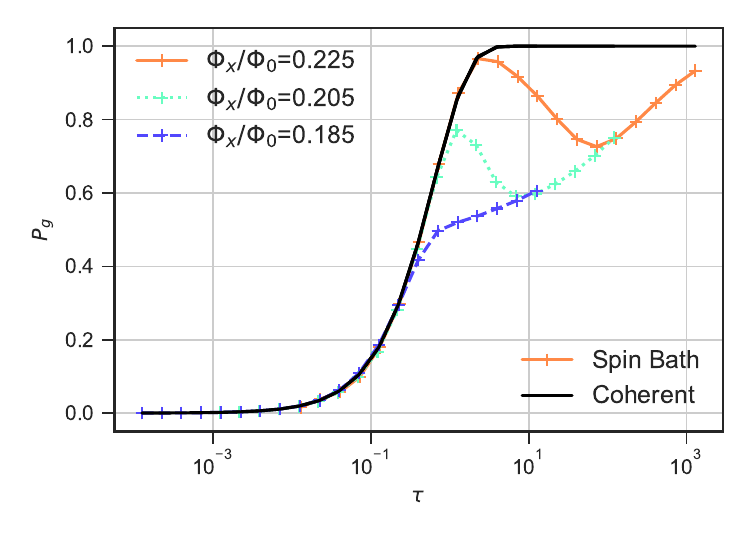}
    \caption[Spin bath simulation results with nominal $1/f$ noise amplitude.]{Simulated LZ final ground state probabilities versus the dimensionless sweep time $\tau$ for different $\Phi_x$ or $\Delta$, using the spin bath model, with 3 spins and targeting the nominal $1/f$ noise amplitude. }
    \label{fig:3SpinNominalAfz}
\end{figure}

%In the main text, 
\bibliographystyle{naturemag}
\bibliography{si}